\documentclass[aps,rmp,preprint,groupedaddress,floatfix]{revtex4-1}
\usepackage{epsfig}
\usepackage{amsfonts}
\usepackage{amsmath}
\usepackage{genetics}

\begin{document}

\newcommand{\mon}{\begin{displaymath}}
\newcommand{\moff}{\end{displaymath}}
\renewcommand{\b}[1]{\mbox{\boldmath ${#1}$}}
\newcommand{\pd}[2]{\frac{\partial {#1}}{\partial {#2}}}
\newcommand{\od}[2]{\frac{d {#1}}{d {#2}}}
\newcommand{\inti}{\int_{-\infty}^{\infty}}
\newcommand{\eon}{\begin{equation}}
\newcommand{\eoff}{\end{equation}}
\newcommand{\e}[1]{\times 10^{#1}}
\newcommand{\chem}[2]{{}^{#2} \mathrm{#1}}
\renewcommand{\sb}{s}
\newcommand{\s}{s}
\newcommand{\eq}[1]{Eq. (\ref{#1})}
\newcommand{\ev}[1]{\langle #1 \rangle}
\newcommand{\mat}[1]{\bf{\mathcal{#1}}}
\newcommand{\fig}[1]{Fig. \ref{#1}}
\newcommand{\degrees}{\,^{\circ}\mathrm{C}}
\renewcommand{\log}{\ln}
\renewcommand{\sec}[1]{section \ref{#1}}
\newcommand{\pr}[1]{P_{#1}}
\newcommand{\like}{\mathcal{L}}

\newcommand{\kav}{k_{\text{av}}}
\newcommand{\defeq}{\equiv}
\newcommand{\ud}{U_d}
\newcommand{\un}{U_n}
\newcommand{\thetad}{\theta_d}
\newcommand{\thetan}{\theta_n}
\renewcommand{\L}{\zeta}
\newcommand{\eaon}{\begin{eqnarray}}
\newcommand{\eaoff}{\end{eqnarray}}

\renewcommand{\baselinestretch}{1.0}

\title{The structure of allelic diversity in the presence of purifying selection}

\author{Michael M. Desai$^{1*}$}
\author{Lauren E. Nicolaisen$^{1*}$}
\author{Aleksandra M. Walczak$^{2}$}
\author{Joshua B. Plotkin$^{3}$}
\affiliation{\mbox{${}^1$Department of Organismic and Evolutionary Biology, Department of Physics, and} \\ \mbox{FAS Center for Systems Biology, Harvard University} \\ ${}^{2}$CNRS-Laboratoire de Physique Th\'eorique de l'\'Ecole Normale Sup\'erieure, \mbox{24 rue Lhomond, 75005 Paris, France,} \\ \mbox{${}^3$Department of Biology, University of Pennsylvania} \\ \mbox{${}^*$These authors contributed equally to this manuscript}}

\begin{abstract}
In the absence of selection, the structure of equilibrium allelic diversity is described by the elegant sampling formula of Ewens.  This formula has helped shape our expectations of empirical patterns of molecular variation.  Along with coalescent theory, it provides statistical techniques for rejecting the null model of neutrality.  However, we still do not fully understand the statistics of the allelic diversity expected in the presence of natural selection.  Earlier work has described the effects of strongly deleterious mutations linked to many neutral sites, and allelic variation in models where offspring fitness is unrelated to parental fitness, but it has proven difficult to understand allelic diversity in the presence of purifying selection at many linked sites.  Here, we study the population genetics of infinitely many perfectly linked sites, some neutral and some deleterious.  Our approach is based on studying the lineage structure within each class of individuals of similar fitness in the deleterious mutation-selection balance. Consistent with previous observations, we find that for moderate and weak selection pressures, the patterns of allelic diversity cannot be described by a neutral model for any choice of the effective population site.  We compute precisely how purifying selection at many linked sites distorts the patterns of allelic diversity, by developing expressions for the likelihood of any configuration of allelic types in a sample analogous to the Ewens sampling formula.
\end{abstract}

\date{\today}

\maketitle

Running Head:  Allelic Diversity and Purifying Selection

Keywords:  Allelic Diversity, Purifying Selection, Ewens Sampling Formula, Linkage

Corresponding Author:

Joshua B. Plotkin

Department of Biology

University of Pennsylvania

219 Lynch Labs

433 S. University Avenue

Philadelphia, PA 19104

215-573-8052

jplotkin@sas.upenn.edu

\clearpage

\newpage

\section{Introduction}

In any evolving population, new clonal lineages are constantly being created and destroyed.  The balance between the creation of lineages by new mutations and their destruction by natural selection and genetic drift determines the statistics of the clonal structure of the population.  In the absence of natural selection, \citet{ewens72} computed an elegant sampling formula describing the clonal structure of a neutral population, and explained how the allelic (i.e. lineage) configuration in a sample of individuals from the population provides a window into this clonal structure.

Natural selection distorts the clonal structure of a population away from this neutral expectation.  Of particular interest is purifying (negative) selection against many linked deleterious mutations (``background selection'').  Recent evidence has suggested this may be generally important in a wide range of populations (see \citet{hahn08} for a recent review).  In this paper, we explore how this type of selection alters the clonal (i.e. allelic) structure of a population.  Our analysis leads to a generalization of the Ewens sampling formula to situations involving background selection.

Over the past few decades, numerous authors have studied allelic diversity in infinite-alleles frameworks that incorporate selection. \citet{li77} and \citet{watterson78} introduced models in which alleles may have a few different selective effects.  \citep{li78} and others \citep{li79, ewensli80, griffiths83} analyzed the structure of allelic diversity in these models.  More recent work has analyzed a very general model of selection introduced by \citet{ethierkurtz87}, which allows for diverse types of selection pressures \citep{ethierkurtz94, joycetavare95, grotespeed02, joyce95}.  This work has helped us understand the general effects of selection in distorting the frequency spectrum of sampled alleles.  However, the models these authors have analyzed cannot be directly connected to a concrete description of mutations and selection occurring at specific sites. Rather, they assume that each new mutation creates a new allele whose fitness is completely independent of the fitness of its parent.  In other words, there is no sense of relatedness among alleles, or of a correlation in fitness between closely related alleles.  \citet{EtheridgePMID19341750} and \citet{EtheridgePMID20685218} have more recently derived a coalescent dual of the Moran process with an arbitrary number of types, mutation rates between types, and genic selection coefficients, but it is not clear how this corresponds to selection acting on some fraction of an infinite number of specific sites.

In this paper we take a different approach, based on the specific model of linked sites described by \citet{kaplanhudson88} and \citet{hudsonkaplan94}.  That is, we imagine that each individual has a genome comprised of many neutral and many negatively selected sites.  The fitness of each individual is determined by the number of mutations it carries at the negatively selected sites.  We make the infinite-sites assumption that no two mutations at the same site ever segregate simultaneously.  This is also an infinite-alleles model, but it is based on a specific model of mutations at individual sites, and the fitness of each new allele depends on the fitness of its parent.

Earlier studies have investigated the effects of purifying selection in models identical or closely related to the one we consider here.   \citet{kaplanhudson88} introduced a model essentially identical to the one we analyze here, and \citet{hudsonkaplan94} developed a simple algorithm which can be used to recursively compute how purifying selection alters the structure of genealogies.  \citet{hudsonkaplan95} and \citet{gordocharlesworth02} further developed this idea, resulting in a simple computational method for sampling genealogical relationships in the presence of background selection.  Related simulation and analytical work has further characterized the structure of genealogies and the statistics of genetic diversity at the level of individual sites in this or closely related models \citep{mcveancharlesworth00, seger10, charlesworth93, comeronkreitman02, comeron08, bartonetheridge04}.  However, this earlier work does not provide an analytic description of lineage structure, or sampling formulae for allelic diversity in the presence of purifying selection on many linked sites.

In this paper, we explicitly analyze the lineage structure, and we derive a selected version of the Ewens sampling formula.  We begin by noting that the balance between mutations at deleterious sites and selection against them leads to a steady state mutation-selection balance \citep{haigh78}.  Our approach is to study the structure of lineages within this steady state, using the Poisson Random Field (PRF) method developed by \citet{sawyerhartl92}.  We show that this lineage structure can alternatively be derived using a retrospective approach, by considering the probabilities of mutation and coalescence events in the ancestry of each individual; these probabilities are calculated by \citet{hudsonkaplan94} and \citet{gordocharlesworth02} (and implicitly in a related context by \citet{bartonetheridge04}).  Our description of lineage structure is thus precisely consistent with the analysis of genealogical structures in this earlier work.  Finally, we use our description of lineage structure to calculate sampling formulae for allelic diversity, and compare our predictions to the results of Monte Carlo simulations.

Provided that selection is strong and deleterious mutation rates are sufficiently small, our results show that the effect of background selection on allelic diversity is to reduce the effective population size without otherwise distorting the lineage structure.  Our results are thus consistent with the effective population size approximation to background selection proposed by \citet{charlesworth93}.  For weaker selection, however, or higher mutation rates, the effective population size approximation breaks down, and the effects of background selection become more complex.  We show that in this case the allelic diversity cannot be described by neutral theory with some appropriately chosen effective population size.  This is consistent with earlier observations that background selection leads to distortions in the structure of genealogies \citep{mcveancharlesworth00, seger10, ofallonadler10, comeronkreitman02, comeron08, bartonetheridge04, gordocharlesworth02}.  Our analysis here allows us to compute precisely how these distortions due to purifying selection at many linked sites alter patterns of allelic diversity, and hence provides an analytical framework for exploring where statistical power may lie to distinguish purifying selection from neutrality.

Our approach relies on the assumption that we can describe the distribution of fitnesses within the population with the steady state mutation-selection balance.  In particular, we neglect fluctuations within this balance.  We analyze the validity of this approximation below.  Related to this approximation, we also neglect the effects of Muller's ratchet.  We discuss this approximation in detail in the Discussion.  We further test the validity of our analysis via Monte Carlo simulations; we find that these approximations are reasonable across a broad parameter regime spanning weak and strong selective pressures.

Our analysis in this paper is limited to allelic diversity, and it does not address the degree of relatedness among sampled alleles.  In other words, our analysis only tells us the probability that individuals are genetically identical, not the distribution of the number of specific sites at which individuals may differ.  Our results are thus not directly comparable to the work described above, which makes predictions about expected diversity at the level of individual sites.  However, while our allele-based results provide an incomplete picture of genetic diversity within the population, they do provide a useful perspective on how purifying selection distorts patterns of molecular evolution.  Most importantly, we are able to make precise analytical predictions about how purifying selection distorts allelic diversity, in ways that cannot be described by a single reduced effective population size.

We would of course like to extend our analysis to predict the expected patterns of variation at the level of individual sites. In \citet{negselcoalescent} we use the framework laid out in this paper as the basis for understanding this more general problem.  This allows us to understand how background selection is related to a time-varying effective population size (and how it differs fundamentally from this) in more quantitative detail than has previously been possible.  However, in this paper we focus exclusively on the more limited analysis of allelic diversity, which provides an essential background for the analysis of the more general problem in \citet{negselcoalescent}.

\section{Model}

We imagine a finite haploid population of constant size $N$.  Each haploid genome
has a large number of sites, which begin in some ancestral state and mutate at
a constant rate.  Each mutation is either neutral or confers some fitness
disadvantage $s$ (where by convention $s > 0$).  We assume an infinite-sites
framework, so there is negligible probability that two mutations segregate
simultaneously at the same site.

We assume that there is no epistasis for fitness, and that each deleterious
mutation carries fitness cost $s$, so that the fitness of an individual with $k$
deleterious mutations is $w_k = (1-s)^k$.  Since we assume that $s \ll 1$, we will
often approximate $w_k$ by $1 - s k$.  Later we comment briefly on extensions to our
method to consider the case when the selection coefficient of a deleterious
mutations is drawn from some fixed distribution.

The population dynamics are assumed to follow the diffusion limit of the standard
Wright-Fisher model.  That is, we assume that deleterious mutations occur at a genome-wide rate $\ud$ per individual per generation (with deleterious mutations assumed to be decoupled from selection).  We define
$\thetad/2 \defeq N \ud$, the per-genome scaled deleterious mutation rate.
Similarly, neutral mutations occur at a rate $\un$ per individual per generation,
and we analogously define $\thetan/2 \defeq N \un$.  We assume that each newly
arising mutation occurs at a site at which there are no other segregating
polymorphisms in the population (the infinite-sites assumption).  Since in this
paper we focus only on allelic diversity, this infinite-sites approximation simply
means that each new mutation creates a unique allele.  Throughout the analysis we
assume that Muller's ratchet can be neglected; we discuss the validity of this
approximation in the Discussion.

We study the case of perfect linkage. In other words, we imagine that all the
sites we are considering are in an asexual genome or within a short enough
distance in a sexual genome that recombination can be entirely neglected.
Although our model is defined for haploids, this assumption means that our
analysis also applies to diploid populations provided that there is no dominance
(i.e. being homozygous for the deleterious mutation carries twice the fitness cost
as being heterozygous).

We believe that this is the simplest possible model based on a concrete picture of
mutations at individual sites that can describe the effects of a large number of
linked negatively selected sites on patterns of genetic variation.  It is essentially equivalent to the model described by \citet{kaplanhudson88} and \citet{hudsonkaplan94}, which has formed the basis for much of the analysis of background selection \citep{charlesworth93, gordocharlesworth02, seger10}.

\section{Analysis}

The balance between mutations and selection leads to a steady state distribution
of fitnesses within the population; this is the well-known `mutation-selection balance'. However, the individuals of a given fitness are not all
genetically homogeneous, but rather comprise a number of different alleles.  The
number and frequency distribution of these alleles depends on how quickly new
alleles are created by deleterious mutations from more-fit individuals, and hence
on the overall fitness distribution.

We begin by describing the relevant aspects of the mutation-selection balance that leads to a steady state distribution of fitnesses within the
population.  Our description of this steady state fitness distribution is entirely deterministic.  Of course, in a finite population,
there will be random fluctuations in the values of $h_k$, the fraction of the population harboring $k$ deleterious mutations. In the most
extreme case, these fluctuations lead to Muller's ratchet.  In our analysis below, we will neglect these fluctuations in $h_k$, assuming that these frequencies are always at their deterministic steady state.  Consistent with this approximation, we will also neglect the effects of Muller's ratchet.  We will then return in a later section to use our results to determine when these approximations are valid.

If we assume for a moment that these approximations are reasonable, we can already guess the form of our result for the allelic diversity.  New alleles are constantly being generated within fitness class $k$ due to deleterious mutations from class $k-1$ and neutral mutations from class $k$.  Within class $k$, all alleles drift \emph{neutrally with respect to each other}.  Therefore, conditional on mutations and selection keeping the frequency of the class at $h_k$, the allelic diversity \emph{within this class} will be the same as in a neutral population of size $N h_k$ in which new alleles are created by mutations at the appropriate rate.  Thus for example the probability two individuals are of the same allelic type is the probability that they are both in the same class $k$ times the appropriate neutral result for the homozygosity within that class, summed over all possible classes.  Sampling formulae for larger samples can be calculated in the analogous way.

The remainder of our analysis in this paper is, essentially, devoted to making this simple intuition precise and showing when it is accurate.  We start by summarizing earlier results for the steady state mutation-selection balance $h_k$, and then compute the allelic diversity in detail, neglecting all fluctuations in $h_k$.  This allows us to see precisely when this approximation is reasonable, and hence prove when the simple intuition described above holds.

\subsection{The steady state fitness distribution}
In our model, all deleterious mutations have the same fitness cost $s$, so we can
characterize individuals by their Hamming class, $k$, relative to the wildtype
(which by definition has $k=0$).  That is, individuals in class $k$ have $k$
deleterious mutations more than the most-fit individuals in the population.  Here
$k$ refers only to the number of \emph{deleterious} mutations an individual has;
individuals with the same $k$ can have different numbers of neutral mutations.  We
normalize fitness such that by definition all individuals in class $k = 0$ have
fitness 1.  Individuals in class $k$ then have fitness $1 - ks$.

Imagine that at a given time a fraction $h_k(t)$ of the population is in class
$k$.  This class is acquiring new individuals due to deleterious mutations arising
in class $k-1$, and it is losing individuals due to deleterious mutations away to
class $k+1$.  It also gains or loses individuals at a rate $- (k-\bar k) s$ due to
selection, where $\bar k$ is the mean $k$ within the population, $\bar k \equiv
\sum k h_k$.  This is illustrated in \fig{fig1}.  Note that the term involving
$\bar k$ simply normalizes the effect of selection (selection favors a class if it
is more fit than the average individual, and vice versa).  This means that on
average $h_k(t)$ will evolve according to the equation \eon \od{h_k(t)}{t} = \ud
h_{k-1} - \ud h_k - (k - \bar k ) h_k s. \label{mutselbaldiffy} \eoff  Note this is a system of $k$
equations for all the $h_k(t)$.  Of course random genetic drift will also affect
the $h_k(t)$, and these deterministic equations are only true on average.  We
return to this point below, but for now we neglect drift and focus on the steady
state distribution.

The steady state fitness distribution (the mutation-selection balance) is given by the values of  $h_k(t)$ after a long time.  We can find this mutation-selection balance by setting the right hand side of \eq{mutselbaldiffy} equal to $0$ for all values of $k$.  This calculation was originally carried out by \citet{kimuramaruyama66} and \citet{haigh78}; they found that the steady state, $\hat h_k$, is given by a Poisson distribution with mean $\frac{\ud}{s}$, \eon \label{mutsel} \hat h_k = \frac{e^{-\ud/s}}{k!} \left( \frac{\ud}{s} \right)^k . \eoff  Note that this means that the average fitness in the population is $1 - \ud$, and that $\bar k = \frac{\ud}{s}$.

\subsection{Allelic diversity within a given fitness class}
\begin{figure}
\centering
\includegraphics[width=6.5in]{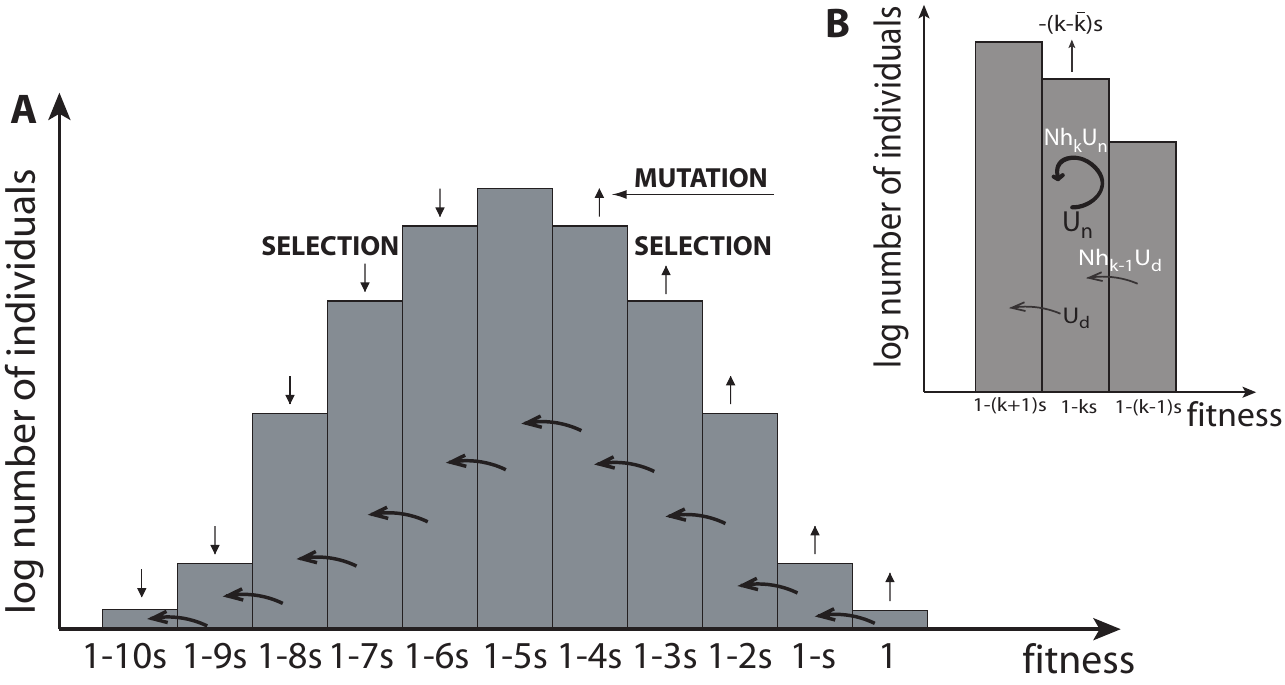}
\caption{\label{fig1} Schematic of the allelic diversity in the mutation-selection balance.  \textbf{(a)} Sketch of the mutation-selection balance in the case $\frac{\ud}{s} = 5$. The steady state distribution of fitness within the population is maintained by a balance between mutations moving individuals towards lower fitness and selection favoring those classes more fit than average at the expense of those less fit than average. \textbf{(b)} The inset shows the processes maintaining a class of individuals with $k$ deleterious mutations. Deleterious mutations from class $k-1$ found new lineages within class $k$ at rate $N h_{k-1} \ud$. Neutral mutations found new lineages in the class at a rate $N h_k \un$.  Selection favors or disfavors individuals from each lineage at a per capita rate $-(k - \bar k) s$, and deleterious mutations eliminate individuals from each lineage at a per capita rate $\ud + \un$.  }
\end{figure}

We now look more closely at individuals within a given fitness class, as illustrated in \fig{fig1}b.  For the moment we neglect neutral mutations; we consider their effects further below.

All lineages in class $k$ originally arose from a deleterious mutation to an individual in class $k-1$. Each of these deleterious mutations founds a new lineage within class $k$.  Such lineages are founded at a rate $\theta_k/2$, where we define \eon \theta_k = 2 N h_{k-1} \ud. \eoff  Note this is true whether or not the $h_k$ are at their steady-state values, though for the purposes of our analysis we will always assume the steady state.

In our infinite-alleles approximation, each new lineage is an allele that is
unique within the population.  The fate of this lineage (allele) is then
determined by the forces of random drift, selection, and additional mutations.
Additional mutations that occur within this lineage go on to found new alleles.
Thus from the point of view of this particular lineage, additional mutations cause
individuals to be lost from the lineage.  This means that individuals are removed
from a lineage in class $k$ at a per capita rate \eon s_k \equiv -\ud - s (k -
\bar k). \eoff  We refer to $s_k$ as the \emph{effective selection coefficient}
against an allele in class $k$, because it is the rate at which any particular
lineage in class $k$ loses individuals (note we have defined signs such that $s_k <0$).  Note that $s_k$ depends implicitly on the $h_k$ through the term involving $\bar k$ (recall $\bar k$ is the average value of $k$, $\bar k \equiv \sum k h_k$).  For convenience we will define the scaled effective selection coefficient $\gamma_k$ by \eon \gamma_k = N s_k. \eoff

Note that in steady state, when the fitness distribution $h_k$ takes the
mutation-selection balance form $\hat h_k$ derived above, $\bar k = \ud/s$ and the
effective selection coefficient $s_k$ is negative for all fitness classes with $k > 0$.  This makes intuitive sense:  each fitness class (except $k=0$) is
constantly receiving new individuals due to mutations.  Thus older individuals
must on average die out, if the fitness class is to stay at a constant steady
state size.  The only exception is the $k=0$ class, for which $s_k = 0$.  This
class drifts effectively neutrally, with its actual selective advantage relative
to the mean exactly balanced by the loss of individuals due to deleterious
mutations.  For $k = 1$ we have $s_1 = -s$, and in general $s_k = - k s$.  On the
other hand, $\theta_k/h_k$ increases with $k$, reflecting the fact that the stronger selection against the larger-$k$ classes is balanced by a larger influx of new deleterious mutations into these classes.

We can now incorporate the effect of neutral mutations.  Each neutral
mutation within an individual in class $k$ creates a new lineage in class $k$.
Thus we may simply redefine the rate at which new lineages are founded, giving \eon
\theta_k \equiv 2 N h_{k-1} \ud + 2 N h_k \un. \eoff  Each neutral mutation also
causes an individual to be lost from the lineage it was in before the mutation, so
we also redefine the effective selection coefficient \eon s_k \equiv -\ud - \un +
s (k - \bar k). \eoff  These neutral mutations are also reflected in \fig{fig1}b.
Note that for all $k$, neutral mutations tend to increase $\theta_k$, and make
$s_k$ more negative.  In the presence of neutral mutations, even $s_0$ is
negative.

We have seen that new lineages are founded within fitness class $k$ at rate $\theta_k/2$, and then drift randomly subject to an effective selective pressure $s_k$.  We now make the key assumption that each lineage is independent of all the others.  This assumption is valid provided that no lineage ever becomes a substantial fraction of the overall population, which will be true whenever $N |s_k| \gg 1$ (i.e. all lineages are selected against strongly enough).  A sufficient condition for this to hold in the bulk of the fitness distribution is simply $N (\un + \ud) \gg 1$, and in fact our approximation will also hold even in some circumstances when this condition breaks down (we describe this further below).

\subsection{Poisson Random Field description of lineage structure}
Using the independence assumption, we have reduced the problem of describing a
lineage within a given fitness class to exactly the situation addressed by the Poisson Random Field model of \citet{sawyerhartl92}. Thus the frequency distribution of lineages (alleles) in fitness class $k$ is a Poisson Random Field (PRF) with parameters $\theta_k$ and $\gamma_k$ (where as before $\gamma_k \equiv N s_k$).  That is, the number of distinct lineages in class $k$ segregating at a frequency between $a$ and $b$ in the entire population is Poisson distributed with mean \eon \int_a^b f_k(x) dx, \eoff where \eon f_k(x) = \frac{\theta_k}{x (1-x)} \frac{1- e^{-2 \gamma_k (1-x)}}{1 - e^{-2 \gamma_k}}. \label{prf} \eoff  This is equivalent to saying that the probability
that there exists a lineage in class $k$ with frequency between $x$ and $x+dx$ is
$f_k(x) dx$, for infinitesimal $dx$.  Note that this PRF result implicitly assumes
that $\theta_k$ and $\gamma_k$ are constant (which requires constant $h_k$), and
hence only describes the diversity in steady state.

This PRF description offers a convenient and well-established way to describe the lineage structure.  It is similar in spirit to the diffusion result used by \citet{ewens72} in his original computation of the neutral ESF.  However, there is an important difference:  Ewens' $f(x)$ was derived as the solution to the diffusion approximation to the $K$-allele Wright-Fisher process, in the limit of infinite alleles.  This explicitly constrains all lineages to add to a total frequency of $1$. The PRF does not impose this constraint.  This makes it possible to compute a simple analytical expression for $f_k(x)$ in the presence of selection.  However, it does involve an implicit approximation.  In Appendix B, we describe this approximation along with a way to relax it using an alternative branching process model to describe lineage structure.

\subsection{The self-consistency condition}
It is clear from our PRF formulation above that the allelic diversity within each
fitness class depends on the $\theta_k$ and $\gamma_k$, which in turn depend on
the $h_k$.  Yet the sum of the frequencies of all the alleles within
fitness class $k$ \emph{is}, by definition, $h_k$.  In steady state, these two
quantities must be equal.  More specifically, we have derived the steady
state value of $h_k$ in \eq{mutsel}, \mon h_k = \frac{e^{-\ud/s}}{k!} \left( \frac{\ud}{s} \right)^k. \moff When we plug these $h_k$ into our PRF result, the summed allele frequencies according to the PRF must agree with steady-state value we used for $h_k$, for consistency.

According to our PRF result, the sum of the frequencies of all the alleles in fitness class $k$ is \eon  h_k = \int_0^1 x f_k(x) \ dx. \label{consistency} \eoff Consistency thus requires that \begin{equation} h_k = \frac{e^{-\ud/s}}{k!} \left( \frac{\ud}{s} \right)^k = \int_0^1 x f_k(x) = \int_0^1 x \cdot \frac{\theta_k}{x(1-x)} \frac{1-e^{-2 \gamma_k (1-x)}}{1 - e^{-2 \gamma_k}} \ dx,   \label{cons1} \end{equation} where $\theta_k$ and $\gamma_k$ depend on $N, \ud, \un, s$ and the $h_k$ as defined above.  Because \eq{mutsel} is equivalent to requiring $\theta_k/2 = |\gamma_k| h_k$ for all $k$ (i.e. in steady state the net influx of individuals into a class must equal the average rate at which individuals within that class are lost), we can rewrite the self-consistency equation as \eon \frac{\theta_k}{2|\gamma_k|}= \int_0^1 x \cdot \frac{\theta_k}{x(1-x)} \frac{1 - e^{-2 \gamma_k(1-x)}}{1-e^{-2 \gamma_k}} \ dx. \eoff  Some algebra reduces this to the condition \eon \int_0^1 \frac{1-e^{-2 \gamma_k x}}{x} \ dx = \frac{1-e^{-2 \gamma_k}}{2 |\gamma_k|}. \eoff  The analysis in Appendix A shows that this condition holds to the level of approximation considered whenever $|\gamma_k| \gg 1$.  When this is true, the steady state mutation-selection balance of \eq{mutsel} is also the distribution $h_k$ that makes our PRF analysis of the allelic diversity within each fitness class self-consistent.

The condition $|\gamma_k| \gg 1$ corresponds to saying that the effective selection coefficient in each class is large compared to $1/N$.  This will be true for all $k$ whenever $N \un \gg 1$.  In practice, even when this condition fails in some fitness classes, it is still valid for all classes in which $|\gamma_k| \gg 1$.  Thus our results still give a good approximation to the population allelic diversity provided $|\gamma_k| \gg 1$ for the classes around $\bar k$ that make up the bulk of the population.  This will hold whenever $\gamma_{\bar k} = N(\ud + \un) \gg 1$.  When this condition does not apply, our PRF result for the allelic diversity within each fitness class is inaccurate.  This is because, when $|\gamma_k | \gg 1$, the growth of some mutant lineages is limited by the size of the population, which is ignored by the PRF approximation. Thus the PRF approximation overestimates the probability that lineages become common, and the self-consistency breaks down.

It is important to note that we also require an additional, stronger condition for other aspects of our analysis to be valid.  The self-consistency condition ensures that the average size of the fitness class implied by the PRF analysis equals the steady state $h_k$.  However, even when this holds, there could be substantial fluctuations in $h_k$ around its average value.  The PRF result for $f_k(x)$ tells us the probability that a set of lineages exists at any given frequencies.  Therefore it contains detailed information about these fluctuations.  However, we have neglected these fluctuations in substituting the $h_k$ into our expressions for $\theta_k$ and $s_k$, and will also neglect these fluctuations below in calculating sampling formulae.  We return to consider this additional approximation in a later section, and also in Appendix B.

\subsection{An alternative, retrospective approach}
It is possible to derive the neutral Ewens sampling formula in two quite different ways.  \citet{ewens72} imagined new alleles being created continuously by new mutations, and considered the frequency distribution of lineages set up by the balance between the continual creation of new alleles and the extinction of older alleles.  This leads to expressions analogous to those in our PRF calculation of the lineage structure.  We can calculate sampling formulas from this lineage structure, as Ewens did in the neutral case.  First, however, we note that in a companion paper to \citet{ewens72}, \citet{karlinmcgregor72} showed that the Ewens sampling formula could also be derived using a retrospective analysis, by considering the ancestral history of a sample of individuals.  This same type of retrospective approach is also possible in our model; in this section we describe this alternative derivation of the allelic diversity as relevant to the case of purifying selection.

In order to calculate the probability of a particular allelic configuration, we consider the ancestral history of a sampled set of individuals.  In particular, we are interested in the most recent event to occur in the history of a sample, backwards in time. We classify these possible events into one of three possible types: coalescence events (i.e. identity by descent), neutral mutations, and deleterious mutations.

This method is easiest to understand if we begin by considering a sample of size two. In order for two individuals to have the same genotype, they of course must be in the same fitness class $k$.  Furthermore, if we look at the ancestral history of each of these two individuals, the most recent event to occur, backwards in time, must be a coalescent event. In contrast, for them to have a different genotype, the most recent event to occur must be a mutation event. Therefore, to calculate the probability of either configuration, we need only calculate the probability that the most recent event is a coalescent event.

In order to calculate the probabilities of each possible most recent event, we must know the distribution of times until each type of event.  In general, neutral mutations are exponentially distributed with rate $U_n$ per generation.  Assuming the steady state values for $h_k$, deleterious mutations are also exponentially distributed with rate $sk$ per generation \citep{hudsonkaplan94}. Finally, within each class, coalescence occurs as a neutral process with rate $\binom{i}{2}$ per $Nh_k$ generations. Therefore, for a sample of size $2$, each of which are sampled from class $k$, we have that: \eaon P(\textrm{1st Event: Coal.})&=&\int_0^{\infty}dt P(\textrm{Coal at t})P(\textrm{No Neut. Mut by t})P(\textrm{No Del. Mut. by t}) \nonumber \\
&=& \int_0^{\infty}dt e^{-t}e^{-2Nh_kU_nt}e^{-2Nh_kskt} \nonumber \\
&=& \frac{1}{1+2Nh_k(U_n+sk)}=\frac{1}{1+\theta_k} ,\eaoff where we have defined $\theta_k\equiv2Nh_k(sk+U_n)$.

This same logic can be easily extended to larger sample sizes.  For example, if we consider $i$ individuals within the same class, the probability that the first event is a coalescence event is  \eaon P(\textrm{1st Event: Coal.})&=&\int_0^{\infty}dt P(\textrm{Coal at t})P(\textrm{No Neut. Mut by t})P(\textrm{No Del. Mut. by t}) \nonumber \\
&=& \int_0^{\infty}dt \binom{i}{2}e^{-\binom{i}{2}t}e^{-iNh_kU_nt}e^{-iNh_kskt} \nonumber \\ &=& \frac{\binom{i}{2}}{\binom{i}{2}+iNh_k(U_n+sk)}=\frac{i-1}{i-1+\theta_k} . \label{retro} \eaoff  If the first event is a coalescence event, that means two of the individuals are of the same allelic type.  This leaves us with $i-1$ individuals in the class which may or may not be identical; we can now use the identical method to ask whether any of these remaining individuals are of the same allelic type.  Similarly, if the first event is a mutation event, the remaining $i-1$ individuals could still coalesce with each other before they also experience mutation events.

We note that our analysis in this section is very similar in spirit to that of \citet{hudsonkaplan94}, \citet{bartonetheridge04}, and particularly to \citet{gordocharlesworth02}.  These earlier authors considered the relative probabilities of mutations and coalescence in the ancestry of each individual, leading to expressions that implicitly contain results analogous to those in this section.  They did not however consider the implications of these results for the overall patterns of allelic diversity in the population, which we now turn to.

\subsection{Sampling formulas}
We can now calculate the probability of sampled configurations of allelic types.  Our goal is to calculate the probability that a sample of $n$ individuals will have some distribution of allelic types (e.g. $n_1$ individuals with allele 1, $n_2$ individuals with allele 2, etc.).  Specifically, we aim to calculate a negative selection version of the neutral Ewens sampling formula (ESF).  As we will see, this calculation proceeds exactly analogously whether we use the lineage structure (PRF) or retrospective analysis.

We begin with the simplest case, a sample of $n=2$ individuals from the
population.  What is the chance that these individuals are the same genotype?  In other words, what is the allelic homozygosity, $Q_2$, in the population?  In order to be
the same genotype, the two individuals must carry the same number of deleterious
mutations --- i.e. they must fall in the same Hamming class, $k$. In addition,
they must also be of the same mutant lineage within class $k$.  This must equal the probability that the most recent event in the history of these $2$ individuals is a coalescence event; from \eq{retro} this is $\frac{1}{1+\theta_k}$.  Alternatively, we could calculate the probability the two individuals are in the same lineage directly from our PRF result; it is the expected value of $x^2$, where $x$ is integrated over
the distribution of lineage frequencies in class $k$: \eon \int_0^1 \frac{x^2}{h_k^2} f_k(x) dx = \frac{1}{1+\theta_k}, \eoff where we have evaluate the integral as described in Appendix A (see also the corrections in Appendix B).

We therefore find that the full probability that two sampled individuals have the same genotype, which we denote $Q_2$, is given by \eon Q_2=\sum_{k=0}^\infty h_k^2\left(\frac{1}{1+\theta_k}\right). \eoff Note that, in the case $U_d=0$, all individuals are in the zero class, such that $h_{k\neq 0}\rightarrow 0$ and $h_0\rightarrow 1$. Therefore: \eon Q_2^{Neutral}\rightarrow\frac{1}{1 + 2NU_n} , \eoff in agreement with the neutral Ewens sampling formula.

In order for two individuals to have a different genotype, there are two possibilities: either the two individuals could be sampled from different classes (in which case they must have a different genotype), or they could be sampled from the same class, and be of different allelic types (cf. the first event in their ancestral history is a mutation event). Therefore: \eaon Q_{1,1} & = & \sum_{k, k'\neq k} h_k h_{k'} + \sum_{k} h_k^2 \left( \frac{\theta_k}{1 + \theta_k} \right) = 1 - \sum_k h_k^2 \left( \frac{1}{1 + \theta_k} \right) = 1 - Q_2 . \eaoff Note that: \eon Q_{1,1}^{Neutral} \rightarrow \frac{2 N U_n}{1 + 2 N U_n} , \eoff in agreement with the neutral Ewens sampling formula.

\subsubsection{Relationship with the Neutral Result}
At this point, it is informative to consider the form of this result. The presence of selection serves to subdivide the population into classes, as given by the mutation-selection balance result. Thus, in order for a sample of individuals to have a particular allelic configuration, they must be sampled from a set of classes consistent with that configuration. However, within each class, the population behaves identically to that of a neutral population, with a different population size ($N\rightarrow Nh_k$) and mutation rate ($U_n\rightarrow U_n+sk$). We can see this explicitly by defining:
\eon Q_{\{Configuration\},k}^{ESF}\equiv \textrm{ESF Result for \{Configuration\} with $\theta\rightarrow 2Nh_k(U_n+sk)$} . \eoff
For example, we have that:
\eon Q_{\{2\},k}^{ESF}=\frac{1}{1+\theta_k} , \qquad Q_{\{1,1\},k}^{ESF} = \frac{\theta_k}{1+\theta_k} . \eoff
We can then rewrite our results as:
\eaon Q_2&=&\sum_{k}h_k^2Q_{\{2\},k}^{ESF} , \\
Q_{1,1}&=&\sum_{k}h_k^2Q_{\{1,1\},k}^{ESF}+\sum_{k,k'\neq k}h_kh_{k'} . \eaoff
Thus we see that, within each class, the probability of a particular configuration is effectively neutral with parameter $\theta=2Nh_k(U_n+sk)$, consistent with our initial intuitive guess for the form of our result. The overall probability of a given allelic configuration is then the probability that a specific configuration is achieved within each class, summed over all possible sets of class configurations that are consistent with the allelic configuration.

\subsubsection{Sample Size $n=3$}
This logic can be extended to larger sample sizes. In order for three randomly-selected individuals to have the same genotype, all three individuals must be sampled from the same class and they must all be from the same lineage (i.e. both of the first two events must be coalescence).  This can be computed by considering the average of $x^3$ over the PRF, $\int_0^1 x^3 f_k(x) dx$, or by using the results from \eq{retro}.  We find: \eon Q_3=\sum_kh_k^3\left(\frac{2}{2+\theta_k}\right)\left(\frac{1}{1+\theta_k}\right) . \eoff
Note that, for $U_d=0$, $h_{k\neq 0}\rightarrow 0$ and $h_0\rightarrow 1$, such that:
\eon Q_3^{Neutral}\rightarrow \frac{2}{(2+\theta)(1+\theta)} , \eoff
in agreement with the neutral Ewens sampling formula.

In order for two individuals to have the same genotype and the third individual to have a different genotype, there are two possibilities. First, two individuals could have been selected from the same class and the third individual could have been selected from a different class. In this case, the two individuals in the same class must be from the same lineage (i.e. coalesce prior to a mutation event). Alternatively, all three individuals could have been selected from the same class. In this case, two must be from the same lineage and the third from a different lineage, which occurs with probability \eon \int_0^1 3 x^2 (1-x) f_k(x) dx.  \eoff  Thinking about this retrospectively, this is equivalent to the sum of two possibilities: either the first event could be a mutation event, in which case the next event among the other two lineages must be a coalescent event, or the first event could be a coalescent event, in which case the next event among the third lineage and the merged lineage must be a mutation event. We find
\eaon Q_{2,1}&=&\sum_{k,k'\neq k}3h_k^2h_{k'}\left(\frac{1}{1+\theta_k}\right)+\sum_k h_k^3\left[ \left(\frac{2}{2+\theta_k}\right)\left(\frac{\theta_k}{1 +\theta_k}\right)+\left(\frac{\theta_k}{2 +\theta_k}\right)\left(\frac{1}{1+\theta_k}\right)\right] \nonumber \\
&=&\sum_{k}\frac{3h_k^2}{1+\theta_k}\left(1-\frac{2h_k}{2+\theta_k}\right) . \eaoff
Note that:
\eon Q_{2,1}^{Neutral}\rightarrow \frac{3\theta}{(1+\theta)(2+\theta)} , \eoff
in agreement with the neutral Ewens sampling formula.

Analogous considerations lead to the probability that all three individuals are of different allelic types, \eaon Q_{1,1,1}&=& \sum_{k,k'\neq k,k'' \neq k',k}h_kh_{k'}h_{k''}+\sum_{k,k'\neq k}3h_k^2h_{k'}\left(\frac{\theta_k}{1+\theta_k}\right)+\sum_kh_k^3\left(\frac{\theta_k}{2+\theta_k}\right)\left(\frac{\theta_k}{1+\theta_k}\right) \nonumber \\ &=&1-\sum_k3h_k^2\left(\frac{1}{1+\theta_k}\right)+\sum_k h_k^3\left(\frac{4}{(1+\theta_k)(2+\theta_k)}\right)=1-Q_3-Q_{2,1} , \eaoff as expected.  Note that \eon Q_{1,1,1}^{Neutral}=\frac{\theta^2}{(1+\theta)(2+\theta)} , \eoff in agreement with the neutral Ewens sampling formula.

\subsubsection{Relationship with the Neutral Result}
As before, we define a class-specific version of the neutral Ewens sampling formula with $\theta\rightarrow 2Nh_k(U_n+sk)$: \eon Q_{\{Configuration\},k}^{ESF}\equiv \textrm{ESF Result for \{Configuration\} with $\theta\rightarrow 2Nh_k(U_n+sk)$} . \eoff In particular, we have that: \eon Q_{\{3\},k}^{ESF}=\frac{2}{(1+\theta_k)(2+\theta_k)}, \qquad Q_{\{2,1\},k}^{ESF}=\frac{3\theta_k}{(1+\theta_k)(2+\theta_k)} , \qquad Q_{\{1,1,1\},k}^{ESF}=\frac{\theta_k^2}{(1+\theta_k)(2+\theta_k)} . \nonumber \eoff Using these formulae, we can rewrite our results: \eaon Q_3&=&\sum_{k}h_k^3Q_{\{3\},k}^{ESF} , \\ Q_{2,1}&=&\sum_{k}h_k^3Q_{\{2,1\},k}^{ESF}+\sum_{k,k'\neq k} 3 h_k^2h_{k'}Q_{\{2\},k}^{ESF} , \\ Q_{1,1,1}&=&\sum_{k}h_k^3Q_{\{1,1,1\},k}^{ESF}+\sum_{k,k'\neq k} 3 h_k^2h_{k'}Q_{\{1,1\},k}^{ESF}+\sum_{k,k'\neq k,k''\neq k',k}h_kh_{k'}h_{k''} . \eaoff Therefore, we again see that, within each class, the probabilities of a particular configuration are effectively neutral with parameter $\theta\rightarrow 2Nh_k(U_n+sk)$. The overall probability of a given allelic configuration is then the probability that a specific configuration is achieved within each class, summed over all possible class configurations that are consistent with the allelic configuration.

\subsubsection{Sampling formulae for arbitrary sample size}
We can extend this method to arbitrary sample size. For example, in order for a sample of $n$ individuals to each have the same genotype, all individuals must be sampled from the same class. They must all be of the same allelic type, which occurs with probability $\int_0^1 x^n f_k(x) dx$.  Or equivalently, the first event among the $n$ lineages must be a coalescent event, the next event among the remaining $n-1$ lineages must also be a coalescent event, and so on. We find
\eaon Q_n&=&\sum_kh_k^n\left(\frac{n-1}{n-1+\theta_k}\right)\left(\frac{n-2}{n-2+\theta_k}\right)\ldots \left(\frac{1}{1+\theta_k}\right) \nonumber \\
& = &\sum_k\frac{h_k^n}{\binom{\theta_k+n-1}{\theta_k}} . \label{qmmaintext} \eaoff
Note that
\eon Q_n^{Neutral}\rightarrow \frac{1}{\binom{\theta+n-1}{\theta}} , \eoff
in agreement with the neutral Ewens sampling formula.

In principle, this method can be extended to calculate the probability of any allelic configuration. Alternatively, we can use the relationship between these results and the neutral Ewens sampling formula to infer the probabilities. We found that, for the cases $n=2$ and $n=3$, we can write the probability of a given allelic configuration as the probability that, within each class, a particular configuration is achieved, summed over all sets of class configurations that are consistent with the allelic configuration. Similarly, we see that for $Q_n$:
\eon Q_n=\sum_kh_k^nQ_{\{n\},k}^{ESF} , \eoff
where we have defined:
\eon Q_{\{Configuration\},k}^{ESF}\equiv \textrm{ESF Result for \{Configuration\} with $\theta\rightarrow 2Nh_k(U_n+sk)$} . \eoff

Using this logic, we can infer the probability of additional configurations. For example, in order for a sample of $n$ individuals of one genotype and $n-m$ of another, there are two possibilities: First, $m$ individuals could be sampled from class $k$ and $n-m$ individuals could be sampled from another class $k'$. The probability of sampling in this manner is $h_k^mh_{k'}^{n-m}\binom{n}{m}$. Within class $k$, the probability of the $m$ individuals having the same genotype is given by the neutral result $Q_{\{m\},k}^{ESF}$ with $\theta\rightarrow 2Nh_k(sk+U_n)$. Similarly, within class $k'$, the probability of the $n-m$ individuals having the same genotype is $Q_{\{n-m\},k'}^{ESF}$. Alternatively, all $n$ individuals could be sampled from the same class $k$. This occurs with probability $h_k^n$. The probability of $m$ individuals having the same genotype and $n-m$ individuals having another is then given by $Q_{\{m,n-m\},k}^{ESF}$. Combining these results and summing over all sets of $k$ and $k'$, we have that:
\eon Q_{m,n-m}=\sum_kh_k^nQ_{\{m,n-m\},k}^{ESF}+\sum_{k,k'\neq k}h_k^mh_{k'}^{n-m}\binom{n}{m}Q_{\{m\},k}^{ESF}Q_{\{n-m\},k'}^{ESF} . \eoff  Note, however, that if $m = n-m$ we must divide by two in the second term in the above expression, to avoid double-counting.

Extending this logic, we have that: \eaon & & Q_{n-m-p,m,p} = \sum_k h_k^n Q_{\{ n-m-p,m,p \}, k}^{ESF} + \sum_{k, k' \neq k} h_k^{n - m - p}h_{k'}^{m + p} \binom{n}{m + p} Q_{\{ n-m-p \}, k}^{ESF} Q_{\{ m,p \},k'}^{ESF} \nonumber \\ & & \; \; + \sum_{k,k'\neq k}h_k^{p} h_{k'}^{n-p} \binom{n}{p} Q_{\{p\},k}^{ESF} Q_{\{ n-m-p,m \},k'}^{ESF} + \sum_{k,k' \neq k} h_k^{m} h_{k'}^{n-m} \binom{n}{m}Q_{\{m\},k}^{ESF} Q_{\{n-m-p,p\},k'}^{ESF} \\ & & \;\; + \sum_{k,k'\neq k,k''\neq k,k'} h_k^{n-m-p} h_{k'}^m h_{k''}^p \binom{n}{n-m-p,m,p} Q_{\{n-m-p\},k}^{ESF} Q_{\{m\},k'}^{ESF} Q_{\{p\},k''}^{ESF} . \nonumber \eaoff Note, however, that we must correct the above expression for overcounting if two or more classes require identical configurations (e.g. if $n - m - p = m = p$ we must divide the second through fourth terms in the above expression by $3$ and the last term by $6$).  In general, the probability of any allelic configuration can be written as the sum over all possible class combinations that are consistent with a given allelic configuration, where the probability of each configuration within a class is given by the neutral result with $\theta \rightarrow 2N h_k (sk + U_n)$.

Note that, in the case $U_d=0$, all individuals are sampled from the zero-class, such that $h_{k\neq 0}\rightarrow 0$ and $h_0\rightarrow 1$. In this case, only the leading-order term will be non-zero in the above results. Therefore, the results reduce exactly to the neutral Ewens sampling formula.

\subsection{Fluctuations in the steady state $h_k$}
Even when the self-consistency condition holds, the frequencies $h_k$ will fluctuate about their steady state frequencies.  However, both our PRF description of the lineage structure and our retrospective analysis assume that the fitness distribution is always in the steady state, $h_k$.  We now consider the validity of this approximation.

Each allele in class $k$ can reach, at most, a maximum size of order $\frac{1}{s_k}$ individuals --- selection prevents any individual allele from becoming more common than this.  The total number of individuals in the class is on average $N h_k$.  This means that, provided that $N h_k \gg \frac{1}{s_k}$, each fitness class is made up of many individual alleles.  Thus we would expect that the fluctuations in the sizes of each one would tend to cancel, and the overall fluctuations in $h_k$ should be negligible provided that this condition holds.  This intuition can be made precise:  we can calculate the variance in $h_k$ in steady state from our PRF approach, or more easily from \eq{heqn} in Appendix B.  We can compute $Var(h_k)/h_k$ from \eq{heqn}, and find that in fact the fluctuations in $h_k$ are indeed negligible provided that \eon N h_k s_k \gg 1. \eoff

In practice, this condition will often not hold in the high-fitness (and low-fitness) tails of the distribution.  However, provided it holds in the center of the fitness distribution from which most individuals will be sampled (i.e. for those fitness classes near the mean), our approach will still give a good approximation to the population allelic diversity.

We note that in addition to assuming $h_k$ are in their steady state values in defining $\theta_k$ and $s_k$ for both the PRF and retrospective approaches, the PRF contains an additional implicit approximation.  In writing the PRF sampling formulae, we assumed that, for example, the probability two individuals in class $k$ come from a lineage of frequency $x$ (given that lineage exists) is $\frac{x^2}{h_k}$.  This implicitly assumes that the the fact that there exists a lineage of frequency $x$ in fitness class $k$ does not impact the expected size of the lineage $h_k$.  That is, we assume that all the lineages in the class always add up to a frequency $h_k$ (i.e., we neglect fluctuations in $h_k$).  This is not strictly true: given that there exists a high-frequency lineage, it is likely that $h_k$ is larger than average, and vice versa.
These correlations between the frequency of an individual lineage and the $h_k$ do not pose a problem to our retrospective analysis, which never makes reference to lineages, but it does lead to small errors in the PRF results.  We show in Appendix B that these errors are negligible provided that fluctuations in $h_k$ can be neglected (i.e. provided $N h_k s_k \gg 1$). However, they do lead to small discrepancies between the PRF and retrospective results (and between the PRF results and the neutral ESF in the $\ud \to 0$ limit, since the neutral ESF is derived assuming a strict constraint on the total population size).  Thus in Appendix B we describe a method to correct for these effects, making the lineage-based and retrospective approaches to allelic diversity exactly equivalent.  All of the above sampling formulae include this correction, as do all our figures.

There is one additional extreme effect of fluctuations in $h_k$:  a fluctuation in $h_0$ can lead to loss of this most-fit class, a process referred to as Muller's ratchet.  We expect that, provided the ratchet does not click many times over the timescale in which individual lineages exist, this will not significantly affect the allelic diversity.  Thus we have neglected the ratchet in our analysis.  We return to consider this in more detail in the Discussion, and test the validity of our approximation with numerical simulations.

\subsection{A distribution of fitness effects of deleterious mutations}
We have analyzed a model in which all deleterious mutations have the same fitness
cost, $s$.  However, in most real populations it is likely that deleterious
mutations have a range of possible fitness effects.  We could model this by
assuming that the overall deleterious mutation rate is still $\ud$, but that
deleterious mutations have a fitness cost between $s$ and $s+ds$ with probability
$\rho(s) ds$.  That is, $\rho(s)$ is the distribution of fitness effects of
deleterious mutations.

In this more general situation, there is still a steady state distribution of
fitness within the population.  Generalizing our earlier notation, we can write
this distribution as $h(k)$, where $N h(k)$ is the steady state number of
individuals with a fitness between $sk$ and $(s+ds)k$, where $s$ is the average
fitness cost of a deleterious mutation and $k$ is no longer constrained to be an
integer.  For certain $\rho(s)$ (e.g. an exponential distribution) it is possible
to calculate $h(k)$ analytically, but even when this is not possible there does
exist some steady state $h(k)$.

The basic ideas behind our analysis still apply in this more general situation.
The rate at which new lineages within fitness ``class'' $h(k)$ are created is now
\eon \theta(k)/2 = N h(k) \un + N \int_0^k h(k') \rho((k-k')/s) dk'. \eoff  The
effective selection pressure against individuals in this class is \eon s(k) = \un
+ \ud - (k-\bar k) s. \eoff  Using these modified parameters, we can now apply our
analysis as before; the distribution of lineage frequencies in class $k$ is given
by the PRF formula $f(k;x)$ with appropriate $\theta(k)$ and $s(k)$.  We can then
find sampling formulas as before --- the only difference is that instead of
summing over a discrete set of fitness classes, we must integrate over a
continuous set of possible fitnesses.  For example, we have $Q_2 = \int_0^\infty
\int_0^1 x^2 f(k,x) dx dk$.

This extension of our model allows us to calculate the effects of more general
forms of purifying selection on allelic diversity.  However, there is a wide
array of possible distributions $\rho(s)$, and using this more general form
obscures the basic effects of selection.  Thus in analyzing our results and
comparing to simulations we focus on the simpler case in which all deleterious
mutations have the same fitness cost $s$.  This focus has the advantage of simplicity,
and it allows us to explore more clearly how the strength of selection affects the
patterns of allelic diversity.

\subsection{Simulations}
In order to check the validity of our analysis, we have performed simulations of a Wright-Fisher population. In our simulations, we consider a
population of constant size $N$ and keep track of the frequencies of all
genotypes over successive, discrete generations.  In each generation,
$N$ individuals are sampled with replacement from the preceding generation,
according to the standard Wright-Fisher process \citep{ewensbook} in which the chance of sampling an individual is determined by its fitness relative to the population mean fitness.

In each generation, a Poisson number of deleterious mutations are
introduced, with mean $N \ud$, and a Poisson number of neutral mutations are
introduced, with mean $N \un$.  The mutations are distributed randomly and
independently among the individuals in the population (so that a single
individual might receive multiple mutations in a given generation).  Each new mutation is ascribed to a novel site, so that each mutation results in a new genotype.

Starting from a monomorphic population, all simulations were run for at least $\frac{1}{s} \log(\ud/s)$ generations (or for at least several times $N$ generations when $\ud/s < 1$), to ensure relaxation both to the steady-state
mutation-selection equilibrium and to the PRF equilibrium of allelic frequencies
within each fitness class.  Appropriate relaxation to steady state was checked by extending the simulations and ensuring our results did not change.  The final state of the population -- i.e. the frequencies of all surviving genotypes -- was recorded at the last generation, and $Q_2$ and $Q_{2,1}$ were calculated from these frequencies.  This was repeated and averaged over 250 replicate simulations to produce the points shown in the figures.

Our simulations allowed for random fluctuations in the frequencies of each fitness class, as well as for Muller's ratchet.  The ratchet did not proceed substantially for the simulations relevant for \fig{fig4}, except for the highest $\ud$ point shown in that figure.  However, it did proceed substantially in the simulations shown in \fig{fig3}, such that the most-fit individuals at the end of each simulation contained typically a few (for small $\ud/s$) to more than a dozen (for larger $\ud/s \sim 10$) deleterious mutations.  We can see that, despite the effects of Muller's ratchet and fluctuations in the $h_k$, our simulations are generally in excellent agreement with our theoretical predictions.

\section{Results and Discussion}

Using the approach we have described, we can calculate the probability of any
allelic configuration within a sample of $n$ individuals from a population
experiencing negative selection at many linked sites.  From this, we can calculate
the expected distribution of any statistic describing allelic diversity.  To do so
we must first determine which allelic configurations lead to what values of the
statistic.  The probability of each possible value of the statistic is then the
sum of the probabilities of all allelic configurations leading to that value.
This is identical to the calculation we would do in the neutral case --- the only
difference is that to calculate the probability of each allelic configuration, we
use our sampling formula rather than the neutral Ewens sampling formula.

In practice, some statistics are easier to calculate than others. While we can easily calculate the distribution of statistics describing diversity in a small sample, and we could in principle calculate certain statistics in larger samples (e.g. the total number of alleles in a sample of size $n$, $K_n$), further work is needed to develop efficient methods of calculating arbitrary statistics in large samples.  This is clearly important for applications of our method to
analysis of sequence data, but the combinatoric and computational issues involved
are an extensive topic which is tangential to the ideas underlying our method.
Instead, we focus here on describing the distributions of simple statistics
involving small samples.  Our aim is to highlight the essential differences
between neutral diversity and the diversity in situations involving linked
deleterious mutations.

\subsection{Relationship to the neutral Ewens sampling formula}
Although it may seem counterintuitive, our analysis applies even when $\ud = 0$ (that is, in the case where all mutations are neutral).  In this case, our model is the same as that studied by \citet{ewens72}.  If we apply our methods to this $\ud = 0$ case, all genotypes are in the fitness class $k = 0$, and we have $h_0 = 1$, $\gamma_0 = - N \un$ and $\theta_0 = \theta = 2 N \un$.  Provided that $|\gamma_0| \gg 1$, the conditions for our PRF analysis to be valid are met, and all of our previous results still apply, but are greatly simplified.  And from our analysis of sampling formulas above we can immediately see that, as expected, setting $\ud = 0$ always causes our results to exactly reduce to the neutral Ewens sampling formula. Note that we must take the limit $\ud \to 0$ rather than $s \to 0$ to recover the neutral result, because taking $s \to 0$ with finite $\ud$ causes the steady state mutation-selection balance to break down (i.e. we have $h_k \to 0$ and fluctuations in the frequencies of each class become crucial).

\begin{figure} \centering \includegraphics[width=3.0in]{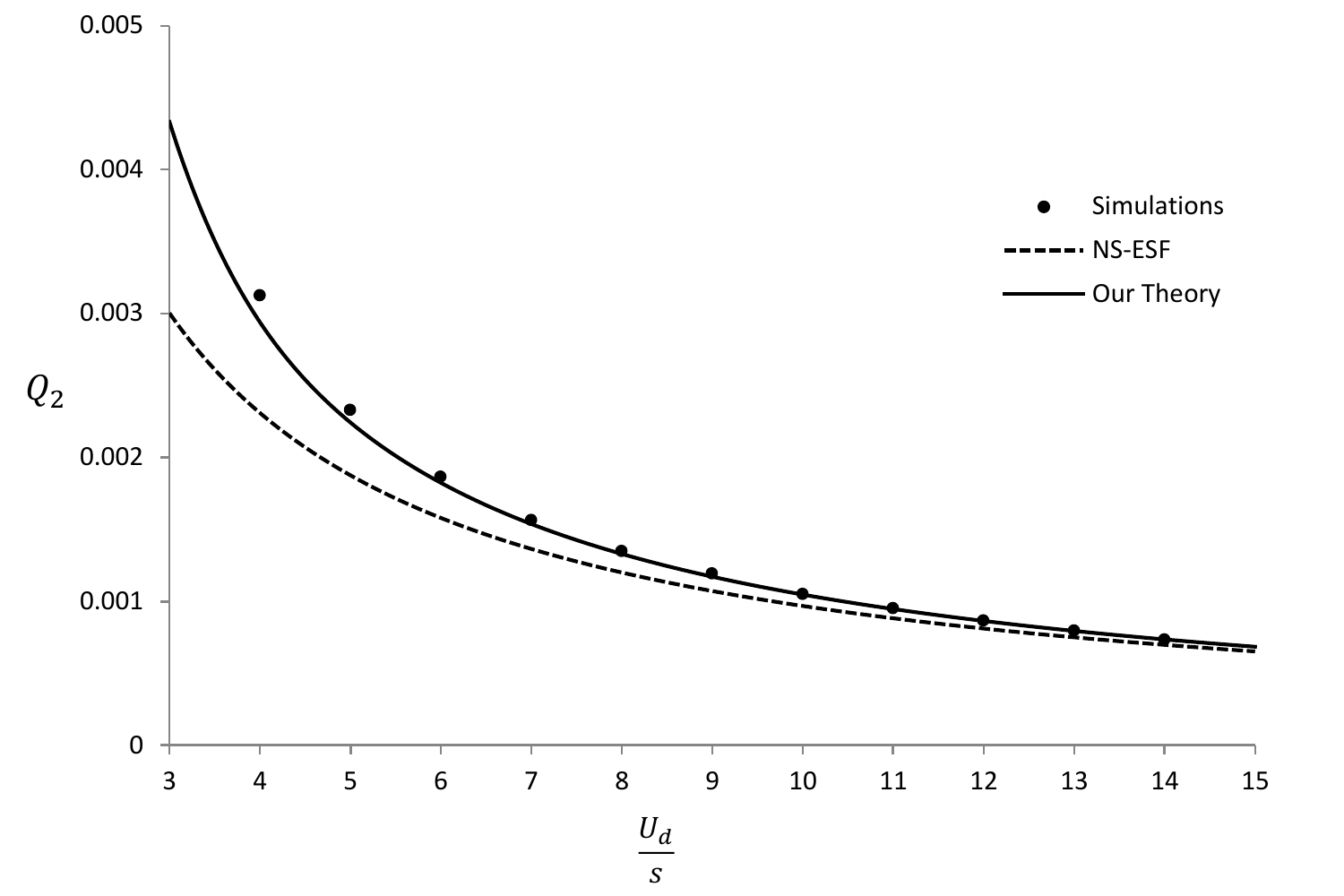}
\includegraphics[width=3.0in]{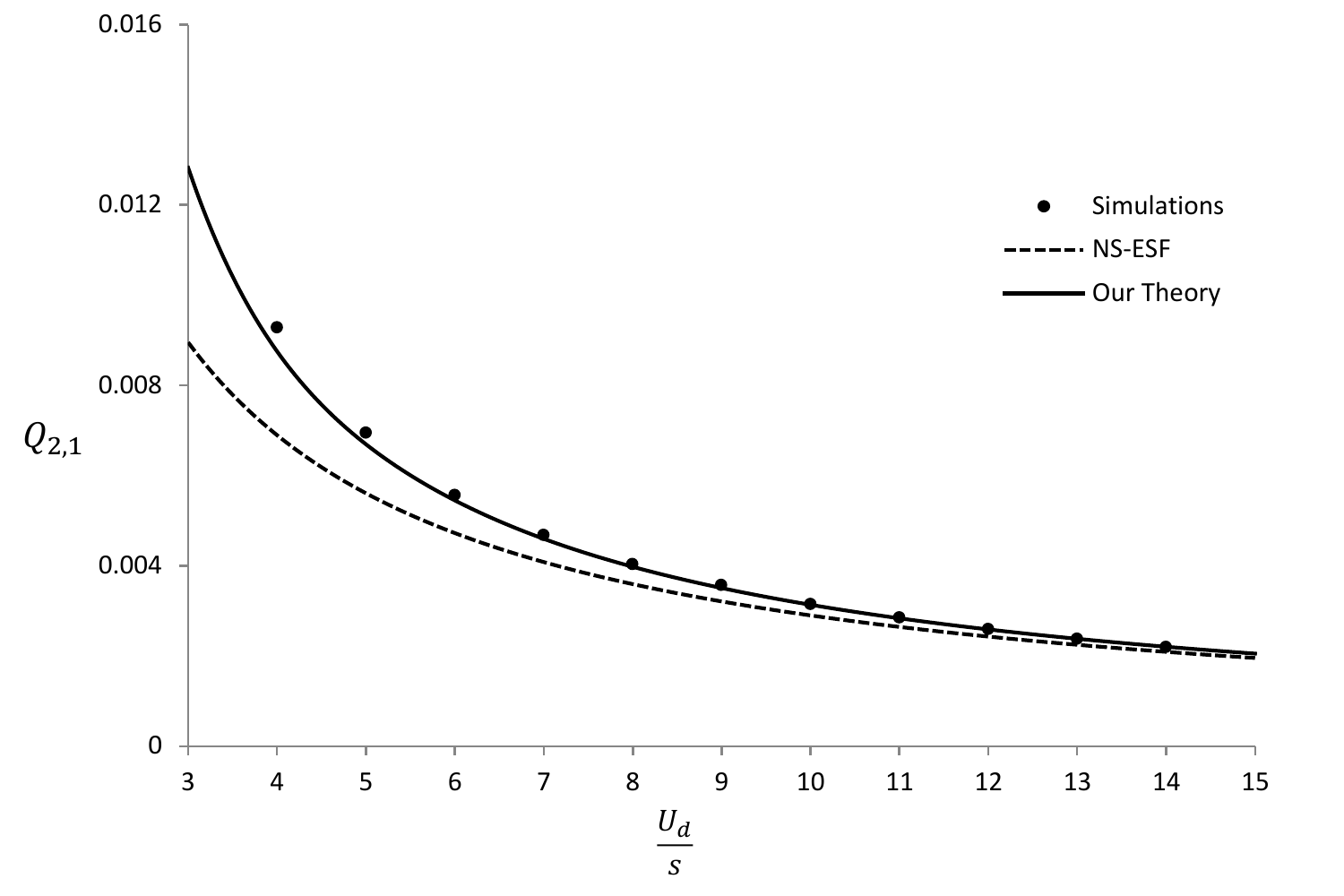}
\includegraphics[width=3.0in]{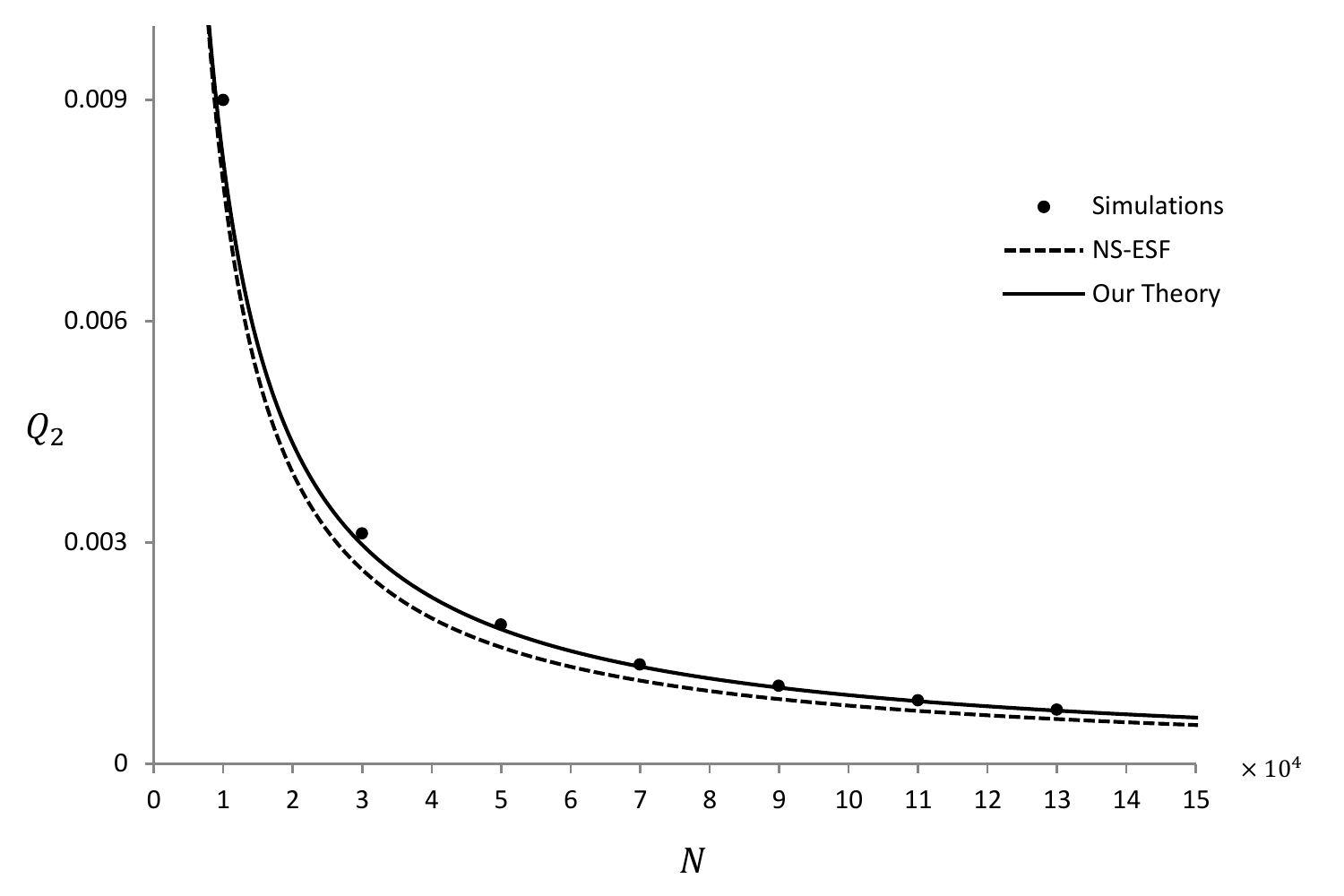}
\includegraphics[width=3.0in]{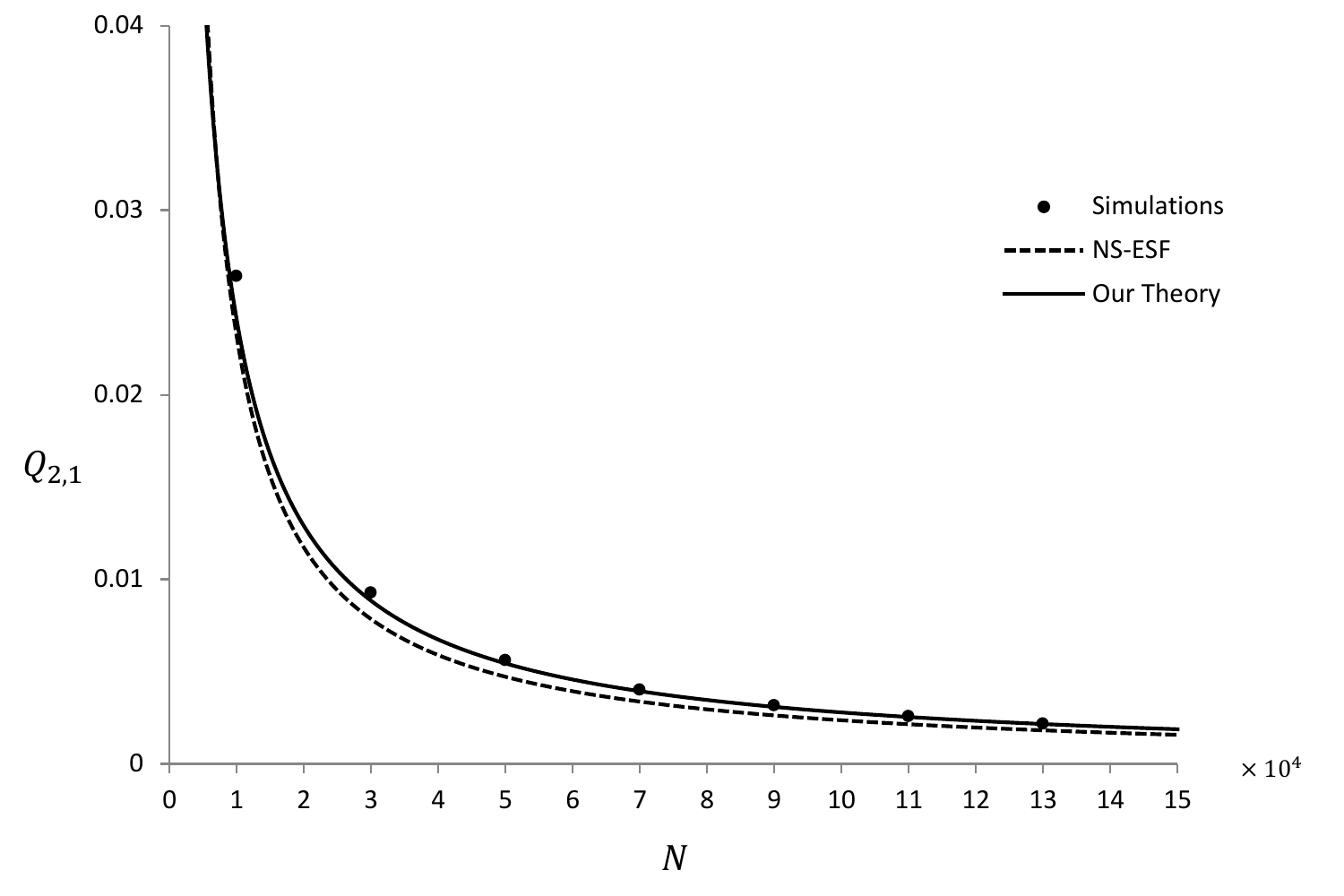}
\caption{\label{fig3}  A comparison between simulation results (dots) and the predictions of our theory (gray lines), for the case where some mutations are deleterious and others are neutral.  For comparison we also show the predictions of NS interpretation of the neutral Ewens Sampling formula (black lines; the NM interpretation gives a worse fit to the data).  \textbf{(a)} Homozygosity $Q_2$ as a function of $\ud/s$ for  $N = 5 \times 10^4$.  \textbf{(b)} $Q_{2,1}$ as a function of $\ud/s$ for  $N = 5 \times 10^4$.   \textbf{(c)} Homozygosity $Q_2$ as a function of $N$ for  $\ud/s = 6$.  \textbf{(d)} $Q_{2,1}$ as a function of $N$ for  $\ud/s = 6$. In all plots  $\un = 3.2 \e{-4}$, $s = 10^{-3}$.}
\end{figure}

For nonzero $U_d$, we expect that our results will differ from the predictions of the neutral ESF.  To illustrate these differences in more detail, we study the allelic configurations in samples of size $n=2$ and $n=3$.  Consider first the homozygosity $Q_2$ in a sample of size $n=2$.  In \fig{fig3}a and c we show how $Q_2$ depends on $\ud$ and the population size $N$, both under our theory and in monte carlo simulations.  We compare these results with the predictions of the neutral ESF. We make the same comparisons for the heterozygosity $Q_{2,1}$ in \fig{fig3}b and d.  We note that the simulation results agree well with our predictions and differ from those of the ESF.

In making this comparison, there is some ambiguity about how to interpret the ESF, which depends only on $\theta$, for $\ud > 0$.  In one interpretation, we neglect selection against the deleterious mutations and set $\theta = 2 N (\un + \ud)$; we refer to this as the NS-ESF case.  Alternatively, we could neglect the deleterious mutations entirely and set $\theta = 2 N \un$; we refer to this as the NM-ESF case.

\begin{figure} \centering \includegraphics[width=5.0in]{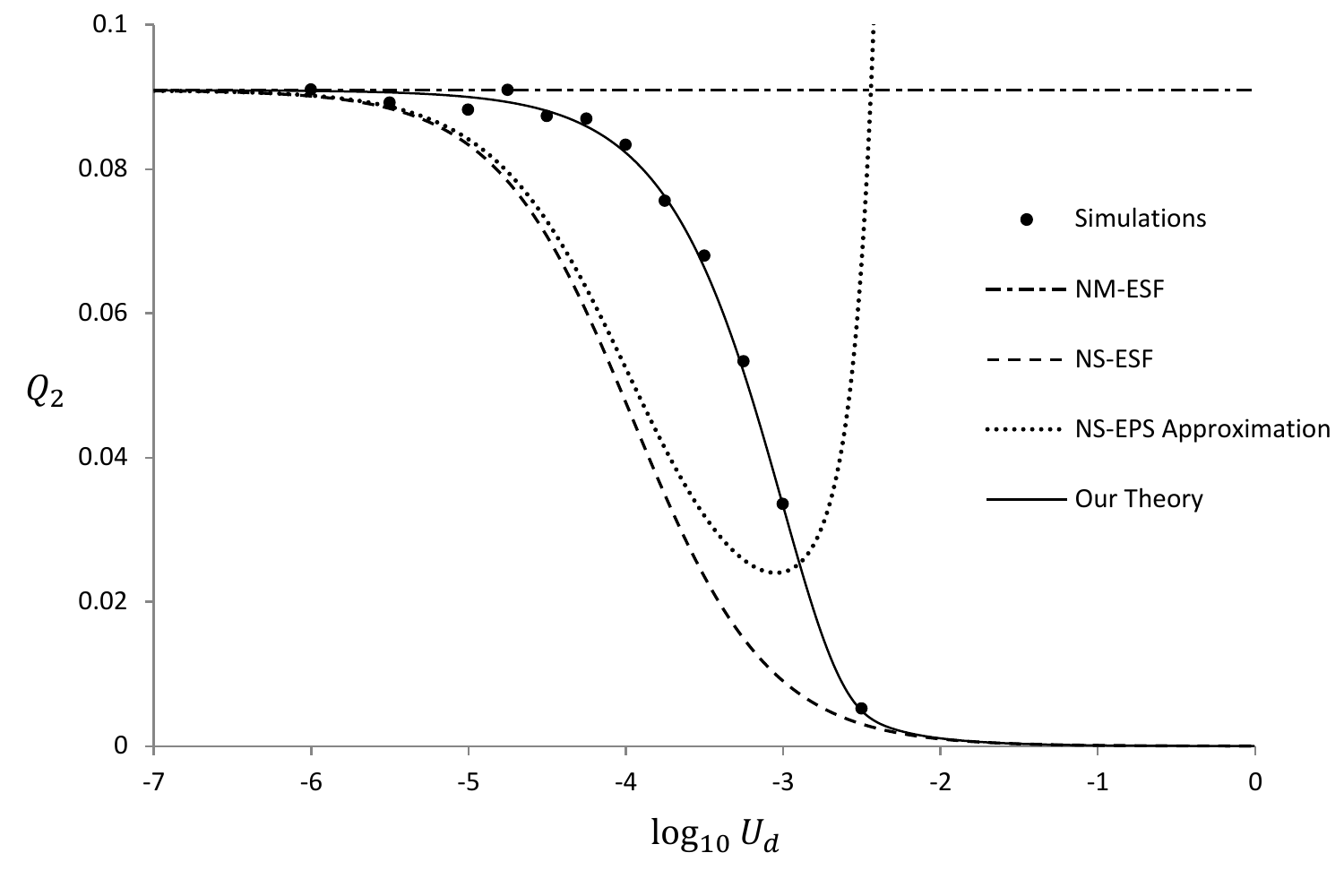}
\includegraphics[width=5.0in]{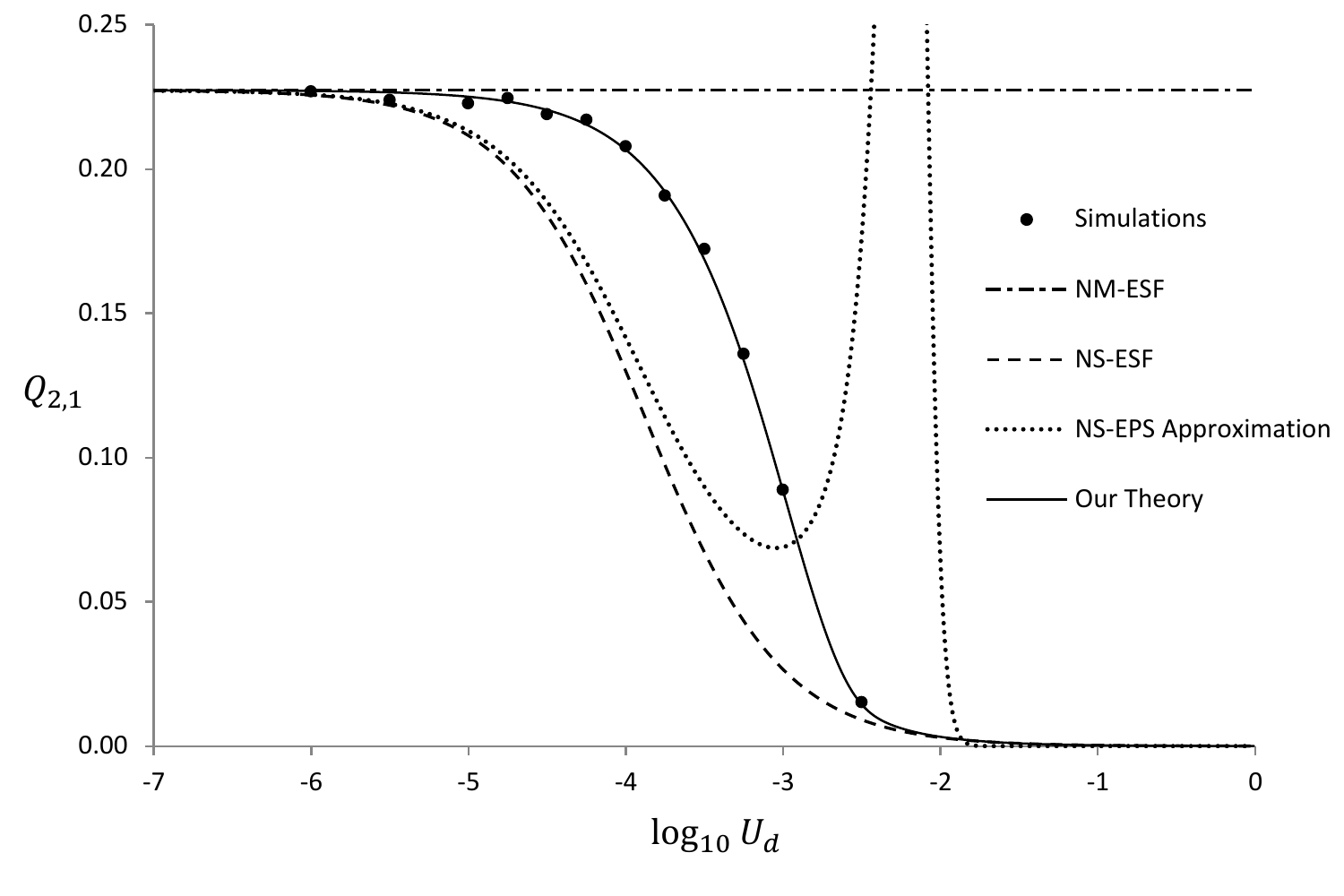}
\caption{\label{fig4}  Allelic diversity as a function of $\log \ud$, for $\un = 10^{-4}$, $s = 10^{-3}$, and $N = 5 \times 10^4$.  Our predictions are shown as a solid line, compared to the predictions of the NS-ESF (dotted line) and NM-ESF (dash-dotted line).  We also compare our results to the predictions of a neutral ESF using the effective population size that would be predicted by background selection (BGS, dashed line), though we emphasize this is not the situation the BGS approximation was developed to address.  These analytical predictions can be compared to simulation results (dots). \textbf{(a)} Homozygosity $Q_2$.  \textbf{(b)} $Q_{2,1}$.  Note that $Q_3 \approx 0$ everywhere for these parameters, so for these predictions $Q_{1,1,1} \approx 1-Q_{2,1}$.  } 
\end{figure}

In \fig{fig4} we explore the ambiguity in the interpretation of the ESF, and compare the predictions of our theory to the two different interpretations of the ESF.  For small $\ud$, our prediction is equivalent to both interpretations of the neutral ESF.  As $U_d$
increases, our predicted homozygosity decreases slowly until it experiences a
sharp transition at $U_d \approx s$.  This transition makes intuitive sense:  when
$U_d < s$, most individuals in the population have no deleterious mutations, and
hence the allelic diversity is similar to the neutral case.  As $U_d$ increases
past $s$, most individuals have deleterious mutations, so these mutations decrease
the expected homozygosity.  These deleterious mutations decrease homozygosity
by less than they would if they were neutral, so our predicted homozygosity is
higher than the NS-ESF but lower than the NM-ESF.

We can gain further insight into this behavior by comparing our predictions to those of the NS-ESF and the NM-ESF in more detail (\fig{fig4}).  We see that even when $\ud = \un$, our predicted homozygosity is only slightly lower than when $\ud = 0$, despite the fact that there are twice as many mutations occurring (and hence the NS-ESF prediction for $Q_2$ has declined by a factor of two).  Here the NM-ESF prediction is fairly accurate, reflecting the fact that selection is still strong (with $\ud \ll s$) so that most individuals have no deleterious mutations at all.  However, as $\ud$ increases past $s$, most individuals now have one or more deleterious mutations and hence these mutations decrease our prediction for the allelic homozygosity.  In this regime, the NM-ESF becomes inaccurate, because the deleterious mutations are sufficiently weakly selected ($\ud \gtrsim s$) that their presence is important to the diversity.  However, despite this being weak selection, the fact that selection eliminates deleterious mutations from the population more rapidly than if they were neutral means that the allelic homozygosity is higher than the NS-ESF, even as $\ud$ becomes very large.  As $\ud$ increases, our predictions become more similar to the NS-ESF, and in the limit of infinite $\ud$ will equal the NS-ESF.  In \fig{fig4}b we show the ``bizygosity'' $Q_{2,1}$ as a function of $\ud$.  Through this parameter range $Q_3$ is small, and so $Q_{1,1,1} \approx 1 - Q_{2,1}$.  As \fig{fig4}b shows, the dependence of bizygosity on $\ud$ is similar to the behavior of heterozygosity, for essentially the same reasons.

This shift in our results from being approximately equal to the NM-ESF for small $\ud$ to the NS-ESF for large $\ud$ has an intuitive explanation from the form of our results for $\theta_k$.  For $\ud \ll s$, $h_0$ is close to $1$, since most individuals have no deleterious mutations.  In this class, we have $\theta_0 = 2 N h_0 s_0 \approx 2 N \un$, the same as the $\theta$ for the NM-ESF.  Since diversity within each class is neutral with the appropriate $\theta$, in this $\ud \ll s$ regime the diversity is approximately that predicted by the NM-ESF.  On the other hand, in the limit of very large $\ud$, $h_k$ becomes sharply peaked about $k = \ud/s$, so almost all individuals have approximately the same fitness, and individual deleterious mutations change fitness by a negligible amount. Thus the diversity is  approximately that predicted by the NS-ESF.  This behavior is exactly as reflected in \fig{fig4}, with the transition between the two regimes occurring at $\ud \sim s$, as this analysis would predict.

Our analysis above makes it clear that the difference between weak and strong selection for the purpose of allelic diversity is set by whether $s$ is small or large compared to $\ud$.  We have potentially three regimes of selection strength.  For $Ns < 1$, selection is ineffective relative to drift, and we always have nearly neutral diversity.  For $Ns > 1$, we can have weak, moderate, or strong selection.  When $s \ll \ud$, we have weak selection as described above; the NS-ESF is accurate.  When $s \lesssim \ud$, we have a ``moderate selection'' regime where the diversity generated by the deleterious mutations themselves can be important, and hence the NM-ESF is inaccurate.  However selection is not so weak that the NS-ESF is accurate either; the selection against the deleterious mutations does reduce the amount of diversity they contribute.  In this regime, neither interpretation of the Ewens neutral sampling formula provides an accurate prediction for allelic diversity. Finally, for $s \gg \ud$, we have a ``strong selection'' regime, where deleterious mutations are eliminated quickly from the population and hence do not contribute to diversity, and the NM-ESF is accurate.  The NS-ESF is also accurate in this regime when $\ud \ll \un$ but it will underestimate homozygosity when $\ud \gtrsim \un$.  Note that in \fig{fig4} we show a case where $s > \un$, so there is a regime where $s \gg \ud$ but $\ud \gtrsim \un$ and hence the NM-ESF is accurate but the NS-ESF is not.  Such a regime does not exist in the case $s < \un$, but otherwise the same qualitative patterns exist for the same reasons.

\subsection{Comparison to the effective population size approximation}
The background selection model we have studied has been the subject of much earlier work, although this has largely been focused on the structure of genealogies in the presence of purifying selection, rather than allelic diversity \citep{hudsonkaplan94, hudsonkaplan95, gordocharlesworth02, seger10}.  A particularly simple and useful approximation to the effects of background selection was developed by \citet{charlesworth93}, \citet{charlesworth94}, and \citet{charlesworth95}.  This approximation is widely used to summarize the effects of background selection \citep{hartlbook}.  We refer to it here as the effective population size approximation (EPS).  The EPS
analysis makes predictions about the the structure of genealogies and hence about genetic diversity at the level of individual
sites, not just the allelic diversity we consider here.  Further, it focuses on
the genetic diversity among neutral mutations only.  Thus it is not directly
comparable to our results in this paper.  Despite this, we find it instructive to
briefly examine how EPS compares to our results, if we apply it to predict allelic
diversity.  We stress that this is not the interpretation intended by
\citet{charlesworth93} and does not provide a fair picture of its accuracy in
general.  Since EPS describes the structure of genealogies, we defer a detailed
discussion of the accuracy of the EPS approximation and its relationship to our
results to \citet{negselcoalescent}, where we calculate the structure of
genealogies under our model.

The EPS approximation assumes that deleterious mutations are eliminated by selection
quickly compared to the coalescence time between two individuals who do not have
any such mutations.  When this is true, almost all neutral mutations we observe
occurred in individuals that did not have any deleterious mutations, because they
have little time to occur in individuals that do have deleterious mutations before
these individuals are eliminated by selection.  Thus, according to the EPS
approximation,
the genetic diversity among neutral sites linked to negatively selected sites
is exactly the same as the entirely neutral case, but with the population size $N$
replaced by the size of the least-loaded (i.e. most-fit) class.  That is, $N$ is replaced by the
effective population size \eon N_e = N h_0 = N e^{-\ud/s}. \eoff  Given this
$N_e$, EPS predicts that any properties of neutral diversity are identical to
those of coalescent theory with the appropriate $N_e$.  Applying this to the
allelic diversity, this predicts that the sampling properties of neutral alleles
will be given by the classical Ewens' sampling formula, using $\theta=2 N \un h_0
= 2 N \un e^{-\ud/|s|}$.  Note this is effectively a NM-EPS case, which seems most
natural.  An alternative NS-EPS case can be defined using $\theta = 2 N (\un +
\ud) h_0$; this leads to similar conclusions.

In the strong selection regime where $\ud \ll s$, most individuals are in the $0$-class. Thus our analysis predicts that this class will dominate allelic diversity, which will be neutral with $\theta_0 = 2 N h_0 s_0 = 2 N e^{-\ud/s} \un$.  Thus our analysis reduces exactly to the predictions of the NM-EPS in this regime.  As we describe in detail in \citet{negselcoalescent}, this is the regime in which the EPS approximation is expected to hold.  Thus we see that our analysis reduces to the EPS in the regime in which it should.

However, for the moderate and weak selection regimes, $\ud \gtrsim s$, the EPS prediction breaks down dramatically.  We graph this prediction in \fig{fig4} (using the NS interpration of the EPS, which provides a slightly better prediction than the NM interpretation).  In this regime the EPS predicts that the neutral homozygosity increases dramatically, since the least-loaded class becomes negligible in size.  However, the homozygosity is not so large in reality, as our predictions demonstrate.  Rather, both neutral and deleterious variation among individuals that harbor one or more deleterious
mutations is important.  Our theory accounts for this effect, while EPS fails
because the approximation that the coalescence time between individuals is
dominated by the time in the least-loaded class breaks down.

We explore the comparison to EPS and the reasons for the breakdown of the EPS
approximation for $\ud \gtrsim s$ in more detail in \citet{negselcoalescent}.
Here, we merely note that, contrary to the intuition one might be tempted to draw
from EPS, having more deleterious mutations can never decrease allelic diversity.
That is, if we fix all other parameters, simply having more deleterious mutations
(i.e. increasing $\ud$) does not reduce heterozygosity.  Certainly it reduces
\emph{neutral} heterozygosity, but accounting for all variation a population with
a larger deleterious mutation rate will have more allelic heterozygosity.

\subsection{Distortions in Allelic Diversity}
The above discussion makes clear that for given population sizes, mutation rates, and selection strengths, purifying selection changes the probabilities of particular allelic configurations in a sample.  However, this does not necessarily imply that selection leads to \emph{distortions} in the patterns of genetic variation compared to the neutral case.  In the neutral case, the probabilities of all allelic configurations in a sample are determined by a single parameter $\theta$.  This means that we can infer $\theta$ from a statistic which depends on the probabilities of one set of allelic configurations, and this $\theta$ then predicts the expected distribution of all other statistics describing genetic variation within the population, provided it is evolving neutrally.

Our discussion of the EPS approximation above makes clear that for sufficiently strong selection, genetic diversity is not distorted relative to the neutral case.  In this section, we show that for moderate to weaker selection (relative to mutation rates), there is no effective population size $N_e$ which can describe genetic diversity in our model. As we noted in the Introduction, this is consistent with earlier observations that background selection leads to distortions in the structure of genealogies \citep{mcveancharlesworth00, seger10, ofallonadler10, comeronkreitman02, comeron08, bartonetheridge04, gordocharlesworth02}.  Here we compute precisely how these distortions alter particular aspects of the patterns of allelic diversity.  Our analysis in this section demonstrates one place in which statistical power exists to distinguish purifying selection from neutral processes at a reduced effective population size.  Our framework can in principle be used to explore where such statistical power lies more generally, but we leave this more general question for future work.

\begin{figure}
\centering
\includegraphics[width=5.0in]{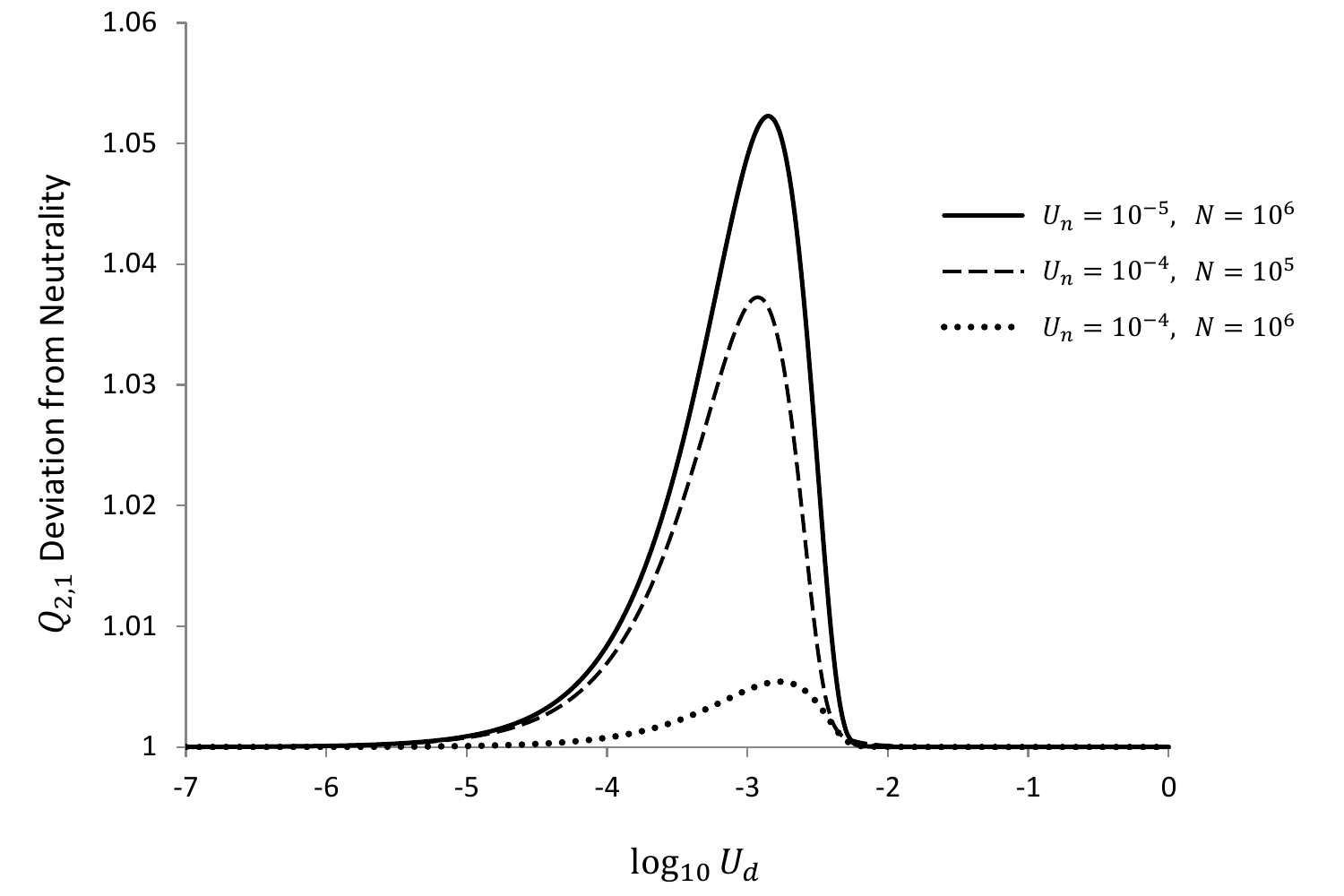}
\includegraphics[width=5.0in]{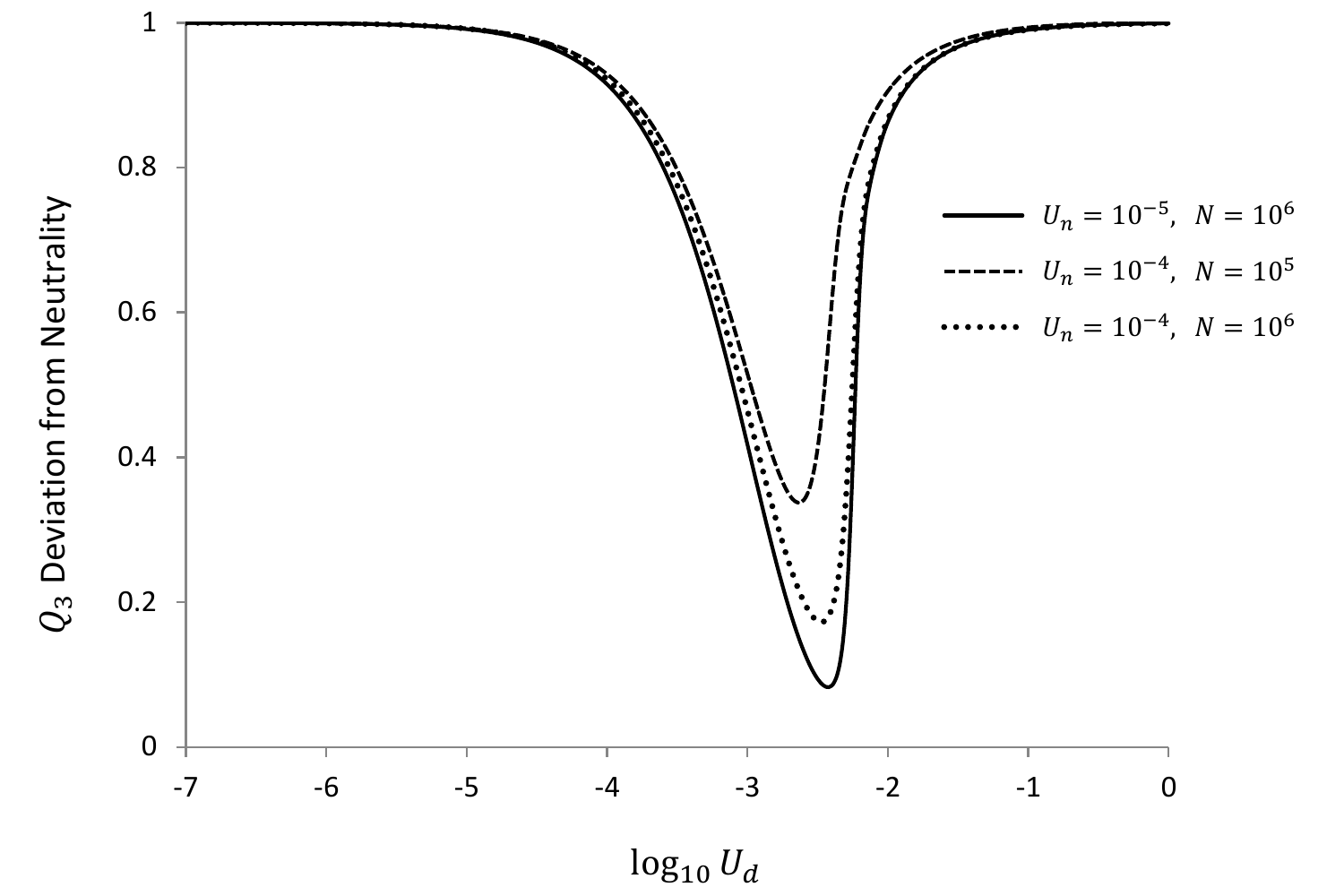}
\caption{\label{fig6}  The deviation from neutrality.  We take $Q_2$ as predicted by our theory, and use the neutral ESF to find the effective $\theta$ that this implies by setting $Q_2 = \frac{1}{1+\theta_e}$.  We then use this effective $\theta_e$ in the neutral ESF to predict the values of $Q_{2,1}$ and $Q_3$ it corresponds to.  We compare this to the $Q_{2,1}$ and $Q_3$ predicted by our theory.  This is a measure of the deviation from neutrality, the skew in the frequency spectrum of allelic diversity away from neutral results with some modified effective population size.  \textbf{(a)} The ratio of $Q_{2,1}$ from the effective population size description to the $Q_{2,1}$ from our theory, as a function of $\log(\ud)$, for $s=10^{-3}$ and three different values of $\un$ and $N$.  \textbf{(b)} The ratio of $Q_{3}$ from the effective population size description to the $Q_{3}$ from our theory as a function of $\log(\ud)$, for $s=10^{-3}$ and three different values of $\un$ and $N$.}
\end{figure}

In this section, we simply show that there is no effective neutral population size $N_e$ to describe diversity in our model.  To do this, it is sufficient to show that the effective $\theta$ that one would infer from one statistic predicts the incorrect values of other statistics.  The simplest way to do this is to begin with the $Q_2$ we would predict given some set of parameters.  We calculate the effective $\theta_e$ one would infer from this $Q_2$ using the neutral ESF (i.e. we choose $\theta_e$ such that $Q_2 = \frac{1}{1 + \theta_e}$).  We then calculate the neutral prediction for $Q_{2,1}$ (or $Q_3$) based on this $\theta_e$.  We compare this with our predictions for $Q_{2,1}$ (or $Q_3$) given the real parameters.  The difference between these two predictions is a measure of the deviation from neutrality.  We show this deviation from neutrality, expressed as the ratio of the neutral effective population size prediction to the actual result, for $Q_{2,1}$ in \fig{fig6}a and for $Q_3$ in \fig{fig6}b.

We see from \fig{fig6} that negative selection distorts the allelic diversity away from high-frequency polymorphisms and towards lower-frequency polymorphisms, for a given level of overall heterozygosity.  The effects are strongest when $\ud$ is of order (or slightly larger than) $s$, and the distortion is stronger for smaller $\un$ and $N$.

These two simple statistics measuring deviations from neutrality demonstrate that there is no effective population size describing allelic diversity.  These particular comparisons are presumably not the most statistically powerful way to detect this type of negative selection, but they do show that statistical power exists.  Using the framework developed in this paper, it is now possible to systematically investigate exactly how linked negatively selected sites generate different patterns of allelic diversity from the neutral case, and to determine which statistics provide the most power detect this type of selection.  Note for example that the deviation from neutrality is much stronger in \fig{fig6}b than in \fig{fig6}a. This reflects the fact that we are inferring $\theta$ from $Q_2$, which in our theory is more closely related to $Q_{2,1}$ than it is to $Q_3$.  Even more powerful tests for selection are presumably possible.  While much earlier work has anticipated that purifying selection distorts the structure of genealogies \citep{mcveancharlesworth00, gordocharlesworth02, hahn08, comeron08, seger10, betancourtcharlesworth09, comeronkreitman02}, no analytic formalism has previously provided a way to determine precisely how selection alters patterns of allelic diversity (and hence, where statistical power may lie).

While we have shown that there is no neutral effective population size describing allelic diversity, this allelic diversity is a summary statistic of the full per-site diversity.  Thus our result also implies that genetic diversity at a per-site level also cannot be described by a neutral effective population size, and that additional power to distinguish neutrality from negative selection can be found in data on site-based variation, consistent with the earlier work described above.  We show how our analysis can be used as the basis for calculating the precise form of this full per-site diversity in a related paper, \citet{negselcoalescent}.

\subsection{Muller's Ratchet}
Throughout our analysis, we have assumed that Muller's ratchet can be neglected.
This is clearly not true in general.  The problem Muller's ratchet creates is that
$h_k$ can change with time, and this changes the distribution of
allele frequencies within each class.  After a ``click'' of the ratchet, the
distribution of $h_k$ shifts, eventually reaching a new state shifted left by one
class (so the class that was originally at frequency $h_k$ is now at frequency
$h_{k-1}$, and so on).  The PRF distribution of lineage frequencies in class $k$
correspondingly shifts from $f_k$ to $f_{k-1}$, and so on, which changes the
allelic diversity.

Fortunately, since $f_k(x)$ is similar to $f_{k+1}$ and $f_{k-1}$, this effect is
unlikely to cause major inaccuracies, provided the ratchet does not click many
times over the timescale on which the lineage frequency spectrum turns over.
We expect that this is generally true within the bulk of the fitness distribution.
At the tails of the distribution, where $h_k$ is small, the allele frequency
distribution can sometimes be substantially different than expected due to the
ratchet.  However, by definition these classes represent a small fraction of the
overall population and hence we do not expect them to contribute substantially to
allelic diversity.

We tested the accuracy of our approximation neglecting Muller's ratchet using the
simulations described above, all of which included the possibility of the ratchet.
Our predictions remain very accurate, even in simulations in which the ratchet was
observed to operate.  Note, however, that the ratchet is potentially more
problematic in considering the genetic diversity at the level of individual sites,
because the high-fitness tail of the fitness distribution can be important for the
structure of genealogies even if it does not contribute substantially to allelic
diversity at any time.  Thus we consider the possible complications introduced by
Muller's ratchet in more detail in \citet{negselcoalescent}.

\subsection{Conclusion}
We have introduced a formalism to calculate the statistics of allelic diversity
in the presence of purifying selection at many linked selected sites.  We have
done so by calculating the structure of the individual lineages that maintain the
deleterious mutation-selection balance.  This analysis is based on the PRF framework of \citet{sawyerhartl92}, which was
originally developed to describe the frequency of mutations at completely unlinked
sites.  We have adapted this framework to our problem with a shift in perspective:
rather than treating new mutations at individual sites as the basic and
independently fluctuating quantities, we consider the lineages founded by new
mutations as the basic independent quantities.  This allows us to describe aspects
of the genetic diversity despite the fact that selection is acting on many linked
non-independent sites.  We showed that this approach is exactly equivalent to a retrospective perspective, which studied the probability individuals are in the same lineage by considering the probability that coalescence events preceded mutations.

Of course, each lineage we describe contains many different mutations, and the fluctuations in lineage frequency described by the PRF framework
represent correlated fluctuations in all of these individual mutations.  If we could also describe how lineages are related to each other, and
hence the statistics of which mutations they share, we could combine this with the results in this paper to describe the full per-site patterns of
genetic diversity despite the correlations between sites introduced by linkage and selection.  We follow precisely this program in a companion paper
\citet{negselcoalescent}.  In this paper, however, we have focused on describing allelic diversity, leading to a negatively selected version of
the neutral Ewens sampling formula.  This analytical framework allows us to compute precisely how patterns of allelic diversity are distorted by
negative selection at many linked sites, and hence understand exactly where statistical power may lie to distinguish purifying selection from
neutrality.

\section*{Acknowledgements}

We thank Warren Ewens and Isabel Gordo for helpful discussions and comments on the manuscript.  MMD acknowledges support from the James S. McDonnell Foundation.  LEN is supported by the Department of Defense through the National Defense Science and Engineering Graduate Fellowship Program, and also acknowledges support from an NSF graduate research fellowship.  Many of the computations in this paper were run on the Odyssey cluster supported by the FAS Sciences Division Research Computing Group at Harvard University.  AMW thanks the Princeton Center for Theoretical Science at Princeton University, where she was a fellow during some of her work on this paper. JBP acknowledges support from the James S. McDonnell Foundation, the Alfred P. Sloan Foundation, the David and Lucille Packard Foundation, the Burroughs Wellcome Fund, Defense Advanced Research Projects Agency (HR0011-05-1-0057), and the US National Institute of Allergy and Infectious Diseases (2U54AI057168).

\newpage

\section*{Appendix A: Integrals involving $f_k(x)$ }

Our expressions for the probabilities of various allelic configurations involve integrals of the form \eon I = \int_0^1 A(x) f(x), \eoff where $A(x)$ is a polynomial function of the form $A(x) = x^n (1-x)^m$ (with $n$ and $m$ integers).  Here $f(x)$ is the expression from \eq{prf}, \eon f(x) = \frac{ah}{e^a-1} \frac{1}{x(1-x)} \left[ e^{a(1-x)} - 1 \right], \eoff  where we have suppressed the subscripts and used the notation $a \equiv - 2 \gamma$.

Whenever $n$ and $m$ are both $\geq 1$, these integrals are easy to evaluate analytically.  When either $n$ or $m$ equals zero, the integrals can be separated into an exactly solvable analytical part and a part that involves the integral \eon I' = \int_0^1 \frac{e^{ay}-1}{y} dy. \eoff  This integral $I'$ is a known special function $Ein(-a)$; see p. 228 of \citet{abramowitzstegun}.

Consider for example the integral \eon I_2 = \int_0^1 x^2 f(x) dx. \eoff  Substituting in for $f(x)$ and substituting $y = 1-x$ in the integral gives \eon I_2 = \frac{ah}{e^a-1} \int_0^1 \frac{1-y}{y} \left[ e^{ay}-1 \right] dy. \eoff  We now simply write $\frac{1-y}{y} = \frac{1}{y} - 1$ and evaluate the analytically solvable parts of this integral to get \eon I_2 = \frac{ah}{e^a-1} I' - h + \frac{ah}{e^a-1}. \eoff

Fortunately, we can calculate a simple analytic approximation for $I'$ in the limit $a \gg 1$ (i.e. $|\gamma| \gg 1$), which is the limit we are always working in.  This is a standard asymptotic expansion of the $Ein$ function; we have \eon I' \approx \frac{1}{a} e^a \left[ 1 + \frac{1}{a} \right] . \eoff  We can now plug our approximation for $I'$ into our result for $I_2$ to get \eon I_2 = \frac{h}{a}. \eoff

For more complex integrals, we need to keep higher order terms in the asymptotic expansion of $I'$.  In general, we find \eon I_n = \int_0^1 x^n f(x) = \frac{(n-1)! h}{a^{n-1}}. \eoff  Similar calculations can be used to find an analogous approximation for $I_m = \int_0^1 (1-x)^m f(x) dx$, but this integral is not necessary for our purposes in this paper.

These calculations allow us to give simple analytic expressions for any integrals of the form $\int x^n (1-x)^m f(x) dx$.  Whenever $m$ and $n$ are both $\geq 1$, the integrals can be evaluated exactly in terms of elementary functions, and when either $m$ or $n$ are $0$ we can use the above results to provide simple analytic approximations to whatever precision we require.

\section*{Appendix B: Fluctuations in $h_k$ and the correlations between the size of a lineage and the frequency of the class }

In the main text, we asserted that our PRF calculations and our retrospective approach were equivalent, and that they reduce precisely to the neutral Ewens sampling formula when $\ud = 0$ and to the EPS approximation when $\ud /s \ll 1$.  However, this is only approximately true: our PRF lineage structure calculations as described thus far are slightly different from the retrospective results (and reduce to the neutral ESF or the EPS approximation to leading order in $\frac{1}{\theta_k}$).  These discrepancies all stem from fluctuations in the size of the fitness classes $h_k$.  In this Appendix, we explain how these discrepancies arise and describe an alternative analysis of lineage structure which allows us to avoid them.

In our PRF calculations for sampling formula in the main text, we wrote that the probability that two individuals picked from class $k$ come from the same lineage is \eon \int_0^1 \left( \frac{x}{h_k} \right)^2 f_k(x) dx. \eoff  The idea behind this equation is that $f_k(x) dx$ is the probability that there exists a lineage in class $k$ at frequency $x$, while $h_k$ is the total frequency of class $k$, so the probability that an individual in class $k$ comes from this lineage (given the lineage exists) is $x/h_k$.  Similarly the probability two randomly chosen individuals come from this lineage is $x^2/h_k^2$, and summing up over all possible lineages (times the probability each exists) gives $\int_0^1 x^2 f_k(x) / h_k^2 dx$.  Analogous expressions were constructed for the probabilities of other allelic configurations.

These expressions assume that the fact that there exists a lineage of frequency $x$ in fitness class $k$ does not impact the expected size of the lineage $h_k$.  This is essentially equivalent to neglecting fluctuations in $h_k$:  we assume that all the lineages in the class always add up to $h_k$.  We saw in the main text that this holds provided that $N h_k s_k \gg 1$.  As we would expect, this is the same condition under which we can neglect fluctuations in the $h_k$.  We have therefore restricted our analysis to this parameter regime.

Even within this parameter regime, this approximation introduces small corrections.  These arise from the fact that, given that there exists a lineage that is at high frequency (a substantial fraction of $h_k$), it is likely that $h_k$ is larger than average.  In other words, there is a correlation between the size of a lineage and the frequency of the class, so the probability that two individuals picked from a class come from the same lineage is not precisely $\int_0^1 \left( \frac{x}{h_k} \right)^2 f_k(x) dx.$  The part of the integrand corresponding to larger $x$ overestimates the probability two individuals are from the same lineage, since given that those high-frequency lineages exist, $h_k$ will be larger than average.  Similarly (though less dramatically), the smaller $x$ part of the integrand underestimates the probability two individuals are from the same lineage.

These problems lead to only slight inaccuracies when our self-consistency equation holds.  However, neglecting these correlations obscures the precise relationship between our results and the neutral Ewens sampling formula (and to our retrospective approach).  Further, relaxing this approximation provides a useful comparison of the subtle differences between the assumptions underlying the original neutral Ewens calculation and the PRF method we use here.  Thus we describe here an alternative approach to understanding the lineage structure in a fitness class which allows us to avoid this approximation.  This eliminates the small discrepancies between our PRF method, our retrospective approach, and (in the $\ud \to 0$ limit) the neutral Ewens formula.

We first note that, in his original calculation of the neutral ESF, \citet{ewens72} used a diffusion result, $f(x)$, roughly analogous to our PRF expression to describe the probability that there exists a lineage with frequency $x$ in the population at a given time.  However, Ewens' $f(x)$ was derived as the solution to the diffusion approximation to the $K$-allele Wright-Fisher process, in the limit of infinite alleles.  This process explicitly imposes the constraint that the sum of all lineages in the population at a given time must add to $1$.  This means that there is no correlation between the size of a lineage and the total number of individuals in the population.

The PRF calculation of the lineage structure does not involve this explicit constraint.  This is what makes it possible to compute a simple analytical expression for $f_k(x)$.  This lack of constraint means that the PRF result admits fluctuations in $h_k$, which lead to corresponding correlations between $x$ and $h_k$.  We could partially avoid this by defining $\gamma_k = N h_k s_k$, rather than $N h_k$, as we have so far.  This would effectively mean that each lineage is assumed to be diffusing between $0$ and $h_k$ rather than between $0$ and $1$, and forbid any lineage from reach a frequency larger than $h_k$.  Thus it reduces the discrepancies associated with the correlations between $x$ and $h_k$.  However, even with this redefinition, there is no constraint that the lineages in a given class all add to precisely $h_k$, and so correlations still exist.

To correct exactly for the effects of correlations between $x$ and $h_k$, we take an alternative approach to the frequency distribution of lineages within a given fitness class.  Rather than use a diffusion approximation to describe the dynamics of each lineage, we use a continuous-time branching process.  As before, we imagine that new lineages are created at a rate $\theta_k/2$.  In steady state there will be some time-independent probability that there are $n$ total individuals across all the lineages in the class, $P(n)$.  Note that on average we must have $n/N = h_k$, and that $P(n)$ contains information on the fluctuations in the $h_k$.  We first compute the generating function for $P(n)$, \eon H(z) \equiv \sum_{n=0}^\infty P(n) z^n. \eoff  To do so, we start by computing the generating function for the probability distribution of the number of individuals from each lineage, as described by Eqs. (7-9) of \citet{desaifisher07}.  We substitute this expression into Eq. (24) of \citet{desaifisher07} and integrate.  We find \eon H(z) \equiv \sum_{n=0}^\infty P(n,t) z^n \equiv \langle z^n \rangle = \left[ \frac{s}{1-z(1-s)} \right]^{\frac{\theta}{2 (1 - s)}}, \label{heqn} \eoff where angle brackets denote expectation values, and we have suppressed the $k$ subscripts.  Note that this equation describes the fluctuations in the size of an individual fitness class:  the mean, variance, and higher moments of $n$ can be easily computed by taking derivatives of $H(z)$.  Note also that this calculation is based on a continuous-time branching process, in which individuals have a different distribution of offspring number than in a Wright-Fisher process, leading to a transient distribution of the frequencies of individual lineages that is half as large as in the Wright-Fisher model for lineages of substantial frequency.  Thus to make comparisons with the Wright-Fisher process, we have to take $\theta \to 2 \theta$ (as we would in comparing Wright-Fisher to Moran models), as described by \citet{desaifisher07}.

We now imagine that there are $B$ sites in the genome, each of which can mutate to create a new lineage in class $k$.  In the large-$B$ limit, each distinct lineage in class $k$ arose from a mutation at a different site in the genome (and we will later make the infinite-sites assumption $B \to \infty$, which makes this exactly true).  The rate at which new mutations found lineages in class $k$ due to mutations at a specific one of these $B$ sites is $\frac{\theta_k}{2B}$.  This means that the generating function for the probability that there are $n$ mutations at a particular site $i$ in class $k$ is \eon H_i(z) = \left[ \frac{s}{1 - z(1-s)} \right]^{\frac{\theta}{B(1-s)}}, \eoff where again we have suppressed the $k$ subscripts and we have taken $\theta \to 2 \theta$ to match to the Wright-Fisher model as described above.

If we define $n_i$ to be the total number of mutants at site $i$ in class $k$, we have that \eon \sigma \equiv \sum_{i=1}^B n_i \eoff is the total number of individuals in the class (note that on average we expect $\sigma = N h_k$).  We now imagine that we sample some number $m$ individuals from class $k$.  The probability that they are all the same allelic type is \eon Q_m^{(k)} = \left\langle \sum_{i=1}^B \frac{n_i^m}{\sigma^m} \right\rangle. \label{qm} \eoff  Note this has the same form as our PRF expression, except we are averaging over $\frac{n_i^m}{\sigma^m}$ rather than averaging over $n_i^m$ and \emph{then} dividing by the average $\sigma^m$.  In other words, we are explicitly accounting for the correlations between $x$ and $h_k$.  Analogous expressions hold for all other possible allelic configurations; we simply take the PRF integrals relevant for that sampling configuration and replace the $h_k$ by $\sigma$, replace the integral over $x$ with taking an expectation, and sum over all $B$ sites.

We now wish to compute these sampling probabilities.  For example, consider $Q_m^{(k)}$.  We can rewrite \eq{qm} using the identity \eon \frac{1}{\sigma^m} = \int_0^\infty \frac{x^{m-1}}{(m-1)!} e^{-x \sigma} dx .  \eoff This identity can easily be verified by integrating the RHS by parts.  Using this, and noting that lineages at each of the $B$ sites are independent, we find \eaon Q_m^{(k)} & = & \left\langle \sum_{i=1}^B n_i^m \int_0^\infty \frac{x^{m-1}}{(m-1)!} e^{-x \sigma} dx \right\rangle \nonumber \\ & = & B \int_0^\infty \frac{x^{m-1}}{(m-1)!} \langle n_1^m e^{-x \sigma} \rangle dx \nonumber \\ & = & B \int_0^\infty \frac{x^{m-1}}{(m-1)!} \langle e^{-x n_i} \rangle^{B-1} \langle n_1^m e^{-x n_1} \rangle dx. \eaoff  The first expectation value inside the integral can be computed by noting that \eon \langle e^{-x n_i} \rangle = H(z=1-x) = \left[ 1 + x \frac{1-s}{s} \right]^{\frac{\theta}{B(1-s)}} . \eoff Differentiating this result $m$ times with respect to $x$ results in an expression for $\langle n_1^m e^{-x n_1} \rangle$.  Plugging these results in and integrating, taking the limit $B \to \infty$, and neglecting higher order terms in $s$, we find \eon Q_m^{(k)} = \theta \sum_{j=0}^{m-1} (-1)^j {m-1 \choose j} \frac{1}{\theta + j} = \frac{(m-1)!}{\prod_{j=1}^{m-1} (\theta+j)} =  \frac{1}{{\theta + m - 1 \choose \theta}}. \eoff

This result is exactly equivalent to the sampling formula described in the main text, \eq{qmmaintext}, and to the result of the retrospective calculation.  It is a correction to the PRF result, \eon Q_m^{(k)} = \int_0^1 x^m f_k(x) dx, \eoff which is identical only to leading order in $1/\theta_k$. Analogous manipulations for other possible sampling configurations show that the method described in this section generally produces sampling formulae that are exactly equivalent to our retrospective method, and which represent second-order corrections to the PRF results.  All of the sampling formulae quoted in the main text and shown in the figures incorporate this correction, which appropriately handles the correlations between the frequency of an individual lineage and the size of the fitness class.  Similarly, in \citet{negselcoalescent}, we also incorporate the correction calculated in this Appendix into our expressions for $I_x^k$ (which is equivalent to the $Q_2^{(k)}$ computed here).

\clearpage

\newpage

\bibliographystyle{genetics}
\bibliography{allelebasedbib}

\clearpage

\newpage

\clearpage

\newpage

\clearpage

\newpage

\clearpage

\newpage

\end{document}